%% file: v1.5_NumericalCoulomb_arXiv.tex
\documentclass[prd,aastex,amsmath,amssymb,twocolumn,floatfix,aps,showpacs]{revtex4-2} 

\usepackage{graphicx} 
\usepackage{amsmath}
\usepackage{amssymb}
\usepackage{textcomp}
\usepackage{xcolor}
\definecolor{myblue}{rgb}{0.0, 0.0, 0.6}
\usepackage{hyperref}
\hypersetup{
  colorlinks = true,
  citecolor  = myblue,
  linkcolor  = myblue,
  urlcolor   = myblue
}

\begin{document}
\title{
  Numerical calculation of Coulomb corrections in forward elastic $\boldsymbol{p}^\uparrow\!\!\;\boldsymbol{p}$ and $\boldsymbol{p}^\uparrow\!\boldsymbol{A}$ scattering
}

\author{A.~A.~Poblaguev}
\email{poblaguev@bnl.gov}
\affiliation{
  Brookhaven National Laboratory, Upton, New York 11973, USA
}

\date{\today}

\begin{abstract}
  The analysis of RHIC hydrogen gas jet target polarimeter measurements of transverse analyzing powers $A_\text{N}(t)$ in proton-nucleus scattering requires accurate Coulomb corrections to both spin-flip and nonflip amplitudes. These corrections must cover a wide range of nuclear charges $Z$ and form factor slopes, with flexibility to vary form factors during data fitting. To avoid technically challenging calculations involving a small but finite fictitious photon mass, the Coulomb correction to the nonflip electromagnetic amplitude with an exponential form factor was related to the corresponding correction for the spin-flip amplitude. This approach allows soft photon contributions to all amplitudes, including those with nonexponential form factors, to be calculated in the massless photon limit using only analytical expressions and numerically stable integrals with nonsingular integrands and finite integration limits. In addition, an absorptive correction to the spin-flip electromagnetic amplitude, which plays a critical role in spin effects in forward polarized proton-nucleus scattering, was accurately evaluated.
\end{abstract}

\maketitle

\section{Introduction}

The Relativistic Heavy-Ion Collider (RHIC) Spin Program \cite{Bunce:2000uv} utilized the Polarized Atomic Hydrogen Gas Jet Target (HJET) \cite{Zelenski:2005mz, Poblaguev:2020qbw} to precisely determine the transverse polarization of proton beams at 100 and 255\,GeV. As part of the polarization measurements, the elastic $pp$ transverse single-spin ($A_\text{N}(t)$) and double-spin ($A_\text{NN}(t)$) analyzing powers were accurately determined in the Coulomb-nuclear interference region (CNI), $0.0013 < -t < 0.018\,\text{GeV}^2$ \cite{Poblaguev:2019saw}.

\begin{figure}[b]
  \begin{center}
    \includegraphics[width=0.85\columnwidth]{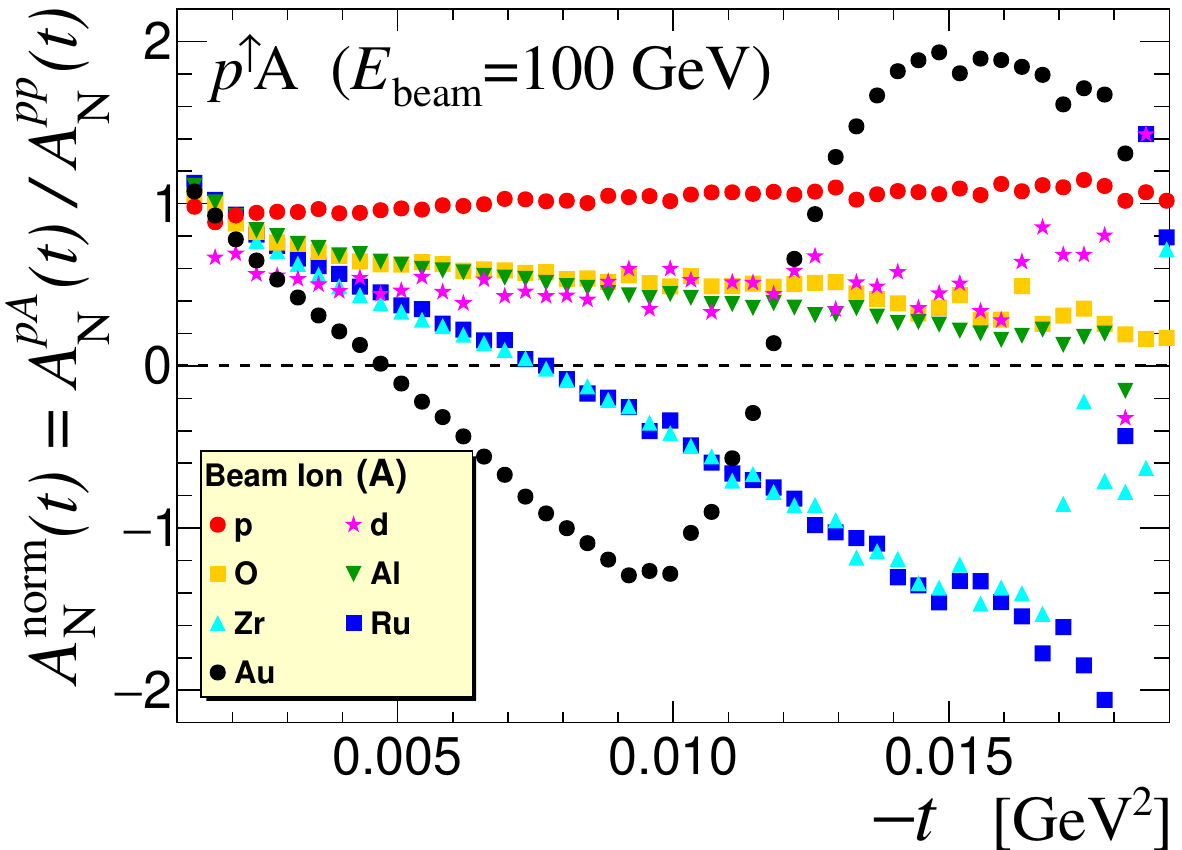}
  \end{center}
  \vspace{-2em}
  \caption{
    Preliminary results for the proton-nucleus elastic analyzing powers $A_\text{N}^{pA}(t)$ measured at HJET\,\cite{Poblaguev:2023stj}. The displayed analyzing powers are normalized to the proton-proton value, calculated for $E_\text{beam} = 100\,\text{GeV}$, assuming no hadronic single spin-flip ($r_5 = 0$).
    \label{fig:pA}
  }
\end{figure}

The good performance of HJET in RHIC heavy-ion beams also enabled experimental evaluation of the proton-nucleus ($p^\uparrow A$) single-spin analyzing power, $A_\text{N}^{pA}(t)$, for several nuclei, as shown in Fig.\,\ref{fig:pA}. These measurements were routinely carried out in a parasitic mode (i.e., without disturbing RHIC experiments) during heavy-ion runs.

In general, the elastic $p^\uparrow A$ transverse analyzing power is given by
\begin{equation}
  A_\text{N}(t) = \frac{2\,\text{Im}\left(F_\text{sf}F_\text{nf}^*\right)}%
  {|F_\text{nf}|^2 + |F_\text{sf}|^2},
  \label{eq:AN}
\end{equation}
where the nonflip (nf) and spin-flip (sf) amplitudes for forward elastic $p^\uparrow A$ scattering can be written\,\cite{Vanzha:1972rps} as
\begin{align}
  F_\text{nf}(\boldsymbol{q}) &= \frac{\sigma_\text{tot}}{4\pi} \left[%
      (i+\rho)e^{-B_Nq^2/2}-\frac{q_c^2}{q^2} e^{-B_Cq^2/2} e^{i\delta_C}
    \right],
  \label{eq:Fnf}
\\
  F_\text{sf}(\boldsymbol{q}) &= \frac{\sigma_\text{tot}}{4\pi}%
  \frac{\boldsymbol{n} \cdot \boldsymbol{q}}{m_p}%
  \left[%
    r_5e^{-B_Sq^2/2}-\frac{\kappa_p}{2}\frac{q_c^2}{q^2} e^{-B_Mq^2/2} e^{i\delta_C}
    \right].
  \label{eq:Fsf}
\end{align}
Here, $\boldsymbol{q}$ is the transverse momentum of the scattered proton, with $q^2 \approx -t$; $q_c^2 = -t_c = 8\pi\alpha Z/\sigma_\text{tot}$, where $Z$ is the nuclear charge; $\boldsymbol{n}$ is a unit vector perpendicular to both the proton beam momentum and spin; and $m_p$ is the proton mass. The parameters $\kappa_p = 1.793$, $\sigma_\text{tot}$, $\rho$, and $r_5$ represent, respectively, the proton anomalous magnetic moment, the total $pA$ cross section, the real-to-imaginary ratio of the hadronic amplitude, and the hadronic spin-flip parameter. The hadronic and electromagnetic form factors are parametrized exponentially with slopes $B_X$, where $X$ denotes one of the indices $N$, $S$, $C$, or $M$. In this work, it is assumed that all slopes $B_X$ are {\em approximately} the same and for the proton-proton  scattering $B_X^{pp}\!=\!8/\left(0.71\,\text{GeV}^2\right)$.

\begin{figure}[t]
  \begin{center}
    \includegraphics[width=\columnwidth]{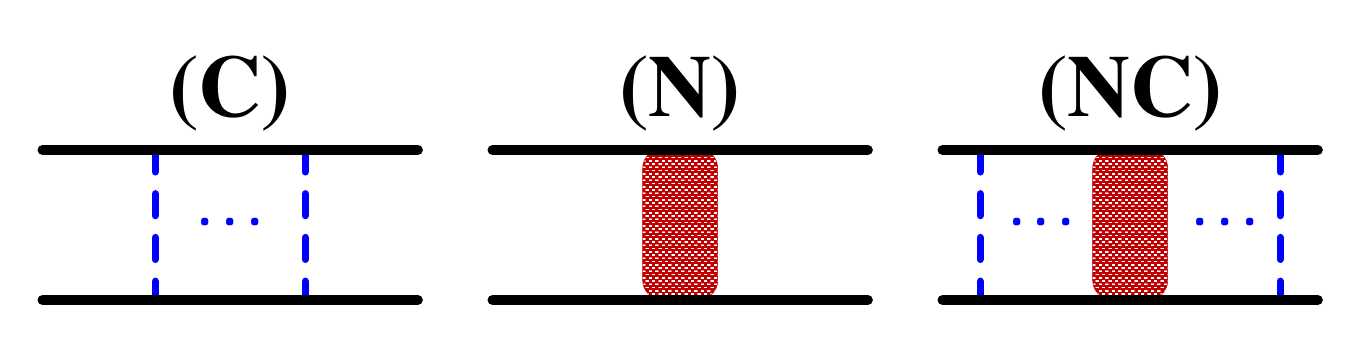}
  \end{center}
  \caption{ \label{fig:graphs}
    Three types of elastic proton-proton scattering: (C) electromagnetic, including multiphoton exchange; (N) bare hadronic; and (NC) combined hadronic and electromagnetic interactions.
  }
\end{figure}

Due to the long-range Coulomb forces (see Fig.\,\ref{fig:graphs}), both the hadronic and electromagnetic (nonflip) amplitudes acquire phase shifts, $\Phi_N(t)$ and $\Phi_C(t)$, respectively. In Eqs.\,\eqref{eq:Fnf} and \eqref{eq:Fsf}, this effect is accounted for by the phase difference\,\cite{Cahn:1982nr}
\begin{align}
    \delta_C(t) &= \Phi_C(t) - \Phi_N(t) \nonumber \\
  &\approx -\alpha Z\left[\gamma - \ln{\frac{(B_N+B_C)|t_c|}{2}} - \ln{\frac{t_c}{t}}\right],
    \label{eq:deltaC}
\end{align}  
known as the Coulomb phase.

For high-energy forward elastic proton-proton scattering, a compact expression for the transverse analyzing power $A_\text{N}(t)$ is well-known\,\cite{Kopeliovich:1974ee,Buttimore:1978ry,Buttimore:1998rj}:
\begin{equation}
  A_\text{N}(t) = \frac{\sqrt{-t}}{m_p}\frac{\left(\kappa_p-2\mathrm{Im}\,r_5\right)t_c/t - 2\mathrm{Re}\,r_5}
  {(t_c/t)^2-2(\rho+\delta_C)t_c/t+1}.
  \label{eq:AN}
\end{equation}
Although some small --- but non-negligible for precise experimental data analysis --- corrections were omitted in Eq.\,\eqref{eq:AN}, the resulting analyzing power is in reasonably good agreement with measurements at RHIC energies.

Structurally, the analyzing power for $p^{\uparrow}A$ scattering is the same as for $p^{\uparrow}p$ scattering. However, due to the nuclear charge $Z$, the much larger Coulomb corrections and the significantly greater absorptive correction to the spin-flip electromagnetic amplitude\,\cite{Kopeliovich:2023xtu} drastically alter the dependence of the proton-nucleus analyzing power on the momentum transfer squared $t$ (as clearly seen in Fig.\,\ref{fig:pA}).

A framework for the theoretical interpretation of experimentally determined proton-nucleus analyzing powers was developed in Ref.\,\cite{Kopeliovich:2023xtu}. In this approach, the hadronic and electromagnetic $pA$ amplitudes are first evaluated using the Glauber model\,\cite{Glauber:1970jm}. Coulomb corrections, including those arising from magnetic photon exchange, are then computed within the eikonal approximation. The calculation proceeds schematically as follows. A Fourier transformation is applied to convert the Born amplitude into the eikonal phase, multiple-photon interactions are accounted for in impact parameter space, and finally, an inverse Fourier transformation returns the result to momentum space.

The primary goal of this paper is to provide clear guidance for the numerical calculation of Coulomb corrections, which are essential for completing the HJET $A_\text{N}^{pA}$ data analysis. The estimates are based on the eikonal approach,\cite{Franco:1973ei,Cahn:1982nr}, and the amplitudes are normalized according to the conventions of Ref.\,\cite{Kopeliovich:2023xtu}. Technical details of the calculations are summarized in the Appendix.

It will be shown that Coulomb corrections to all relevant $p^{\uparrow}p$ and $p^{\uparrow}A$ amplitudes --- including electromagnetic and hadronic, nonflip and spin-flip --- can be computed with the required accuracy without introducing a fictitious photon mass. In addition, the effect of the soft magnetic photon correction to the hadronic nonflip amplitude --- also interpretable as an absorptive correction to the electromagnetic spin-flip amplitude\,\cite{Kopeliovich:2023xtu} --- will be accurately evaluated.

\section{Nonflip proton-proton scattering \label{sec:pp_nf} }

In the eikonal approach used in Ref.\,\cite{Kopeliovich:2023xtu}, the elastic nonflip $\mathit{pp}$ amplitude is expressed via the profile function $i\Gamma(b)$ in impact parameter space as
\begin{align}
  F_\text{nf}(q) &= \frac{1}{2\pi}%
  \int{d^2\boldsymbol{b}\, e^{-i\boldsymbol{q}\!\cdot\!\boldsymbol{b}}\, i\Gamma(b)}
  \nonumber \\
  &= \int_0^\infty{b\, db\, i\left[1-e^{i\chi_N(b)+i\chi_C(b)}\right] J_0(qb)}.
\end{align}
Using the notation
\begin{equation} 
  \gamma_C(b) = i\left(1 - e^{i\chi_C(b)}\right),\quad%
  \gamma_N(b) = i\left(1 - e^{i\chi_N(b)}\right),    
\end{equation}
the profile function can be written as
\begin{align} 
    i\Gamma(b) 
      &= \gamma_C(b) + \gamma_N(b)\, e^{i\chi_C(b)}  \\
      &= \gamma_C(b) + \gamma_N(b) + i\gamma_N(b)\gamma_C(b),
\end{align}
representing a sum of three terms corresponding to graphs (C), (N), and (NC), respectively, in Fig.\,\ref{fig:graphs}.

The eikonal phases $\chi_C(b)$ and $\chi_N(b)$ are Fourier transforms of the corresponding electromagnetic and hadronic amplitudes in the Born approximation:
\begin{equation}
  f_C(q^2) = \frac{-2\alpha}{q^2+\lambda^2}e^{-B_C q^2/2},
  \label{eq:fqC}
\end{equation}
where a small fictitious photon mass $\lambda$ is introduced to regularize the Fourier integral, and
\begin{equation}
  f_N(q^2) = (i+\rho)\frac{\sigma_\text{tot}}{4\pi}e^{-B_N q^2/2}.
  \label{eq:fqN}
\end{equation}

Using Eqs.\,\eqref{eq:chiC-10} and \eqref{eq:fbN}, the eikonal phases can be written in analytic form:
\begin{align}
  \chi_C(b) &= -2\alpha\, \hat{\chi}_C\left(b\sqrt{2/B_C}, B_C\lambda^2/2\right), \\
  \gamma_N(b) &= (i+\rho)\frac{\sigma_\text{tot}}{4\pi B_N}e^{-b^2/2B_N}.
  \label{eq:gammaN}
\end{align}
Since the Fourier transformation is applied directly to the experimentally determined hadronic amplitude \eqref{eq:fqN}, the result in Eq.\,\eqref{eq:gammaN} is the eikonal amplitude $\gamma_N(b)$, rather than the eikonal phase $\chi_N(b)$.

\begin{figure}[t]
  \begin{center}
    \includegraphics[width=0.85\columnwidth]{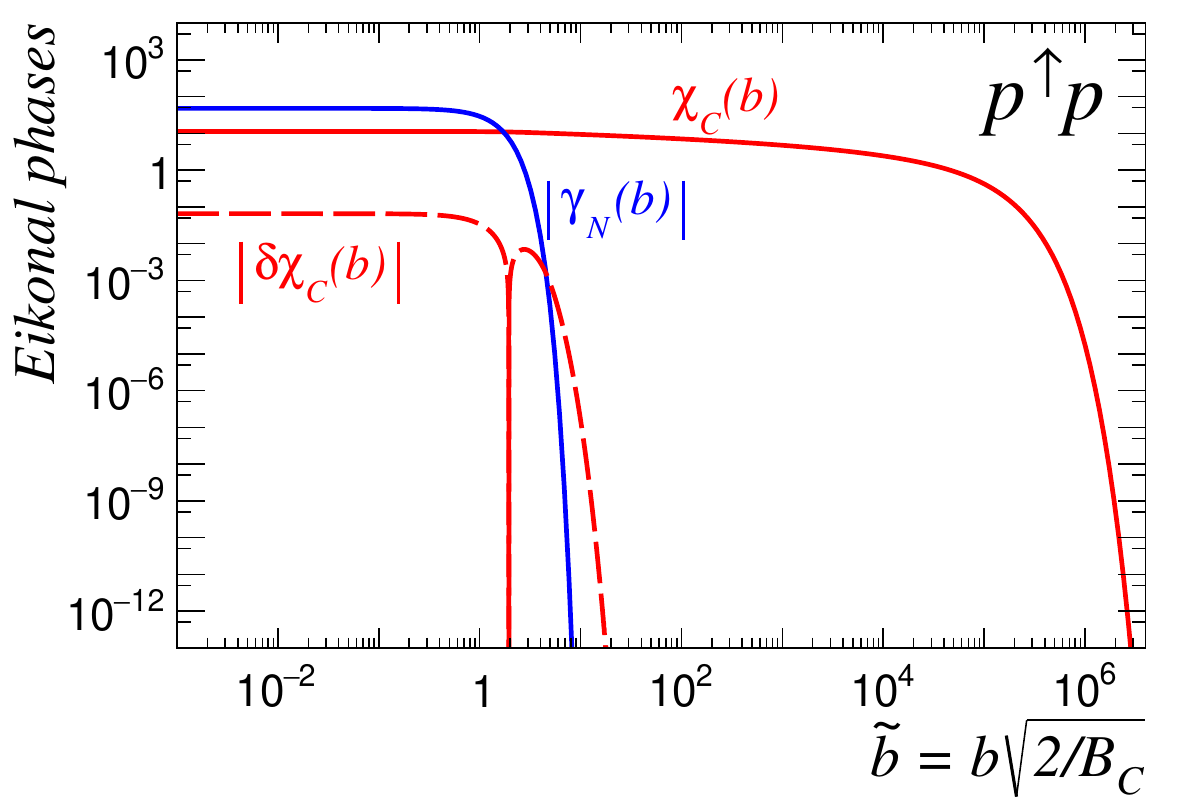}
  \end{center}
  \vspace{-2em}
  \caption{
  Dependence of the eikonal function on the impact parameter. The Coulomb eikonal, $\chi_C(b)$, is normalized according to Eq.\,\eqref{eq:fbC}. The relative normalization of the other functions corresponds to $pp$ scattering with $B_X = 11.3\,\text{GeV}^{-2}$ and $q_c^2 = 1.9 \times 10^{-3}\,\text{GeV}^2$.
  \label{fig:fb}
}
\end{figure}

Relatively normalized functions $\chi_C(b)$ (for $\tilde{\lambda}^2\!=\!10^{-10}$) and $\gamma_N(b)$ are shown in Fig.\,\ref{fig:fb}. Notably, $\gamma_N(b)$ is significant only within a limited range $\tilde{b}\!\lesssim\!10$, while the range of $\chi_C(b)$, $<\!40/\tilde{\lambda}$, is inversely proportional to the photon mass and may be very large.

\subsection{Coulomb correction in case of exponential form factor}

The Coulomb-corrected amplitudes, $f_C^\gamma(q^2)$ and  $f_N^\gamma(q^2)$, can be calculated via the inverse Fourier transform of $\gamma_C(b)$ and $\gamma_N(b)\exp{(i\chi_C(b)}$, respectively. Following the explanation in Appendix\,\ref{sec:chiC'},
\begin{align}
  f_C^\gamma(q^2) &= e^{-i\Delta} \int_0^\infty b\,db\,i\left[1-e^{\chi_C(b)}\right]J_0(bq),
  \nonumber \\
  &=  f_C(q^2)\times\mathcal{F}_C(q^2),
  \label{eq:FgC} \\
  f_N^\gamma(q^2) &= \int_0^\infty b\,db\,\gamma_N(b)\,e^{i\chi'_C(b)}J_0(bq),
  \nonumber \\
  &=  f_N(q^2)\times\mathcal{F}_N(q^2),
  \label{eq:FgN}
\end{align}
where $\Delta$ and $\chi'_C(b)$ are defined in Eqs.\,\eqref{eq:Delta} and \eqref{eq:chiC'}, respectively.

For $\mathit{pp}$ scattering, Coulomb corrections can be approximated by Coulomb phases:
\begin{equation}
  \mathcal{F}_C(q^2)\approx e^{i\Phi_C(q^2)},\qquad%
  \mathcal{F}_N(q^2)\approx e^{i\Phi_N(q^2)}.
\end{equation}
In Ref.\,\cite{Kopeliovich:2000ez}, these phases were calculated analytically to leading order in $\alpha$. After adjusting for the phase shift $\Delta$ used in the present context, the results of \cite{Kopeliovich:2000ez} can be written as
\begin{align}
  \Phi_{N}^\mathrm{KT}(q^2)/\alpha &= \ln{\tilde{q}^2} +\ln{\beta^2}%
   -\mathrm{Ei}\left(\tilde{q}^2\frac{\beta^2}{1+\beta}\right), \\ 
  \Phi_{C}^\mathrm{KT}(q^2)/\alpha &= \ln{\tilde{q}^2} 
  -\mathrm{Ei}\left(\tilde{q}^2/2\right)
  \nonumber \\&\qquad+
  e^{\tilde{q}^2}\left[
    2E_1\left(\tilde{q}^2\right) - E_1\left(\tilde{q}^2/2\right)
    \right], 
\end{align}
where $\tilde{q}^2\!=\!B_Cq^2/2$ and $\beta\!=\!B_N/B_C$.

Since $J_0(x)$ is an oscillatory function with a half-period of approximately $\pi$, accurate numerical evaluation of the integral in Eq.\,\eqref{eq:FgC} may require billions of sampling points (for $\tilde{\lambda}^2 = 10^{-10}$), making the computation time consuming. However, $\mathcal{F}_C(\tilde{q}^2)$, expressed in terms of the dimensionless variable $\tilde{q}$, is a parameter-free function aside from its dependence on the arbitrarily chosen photon mass $\tilde{\lambda}^2$. Therefore, for any fixed value of $\tilde{\lambda}^2$, it is sufficient to tabulate $\mathcal{F}_C(\tilde{q}^2)$ once, with the desired precision, to accelerate subsequent calculations.

\begin{figure}[t]
  \begin{center}
    \includegraphics[width=0.85\columnwidth]{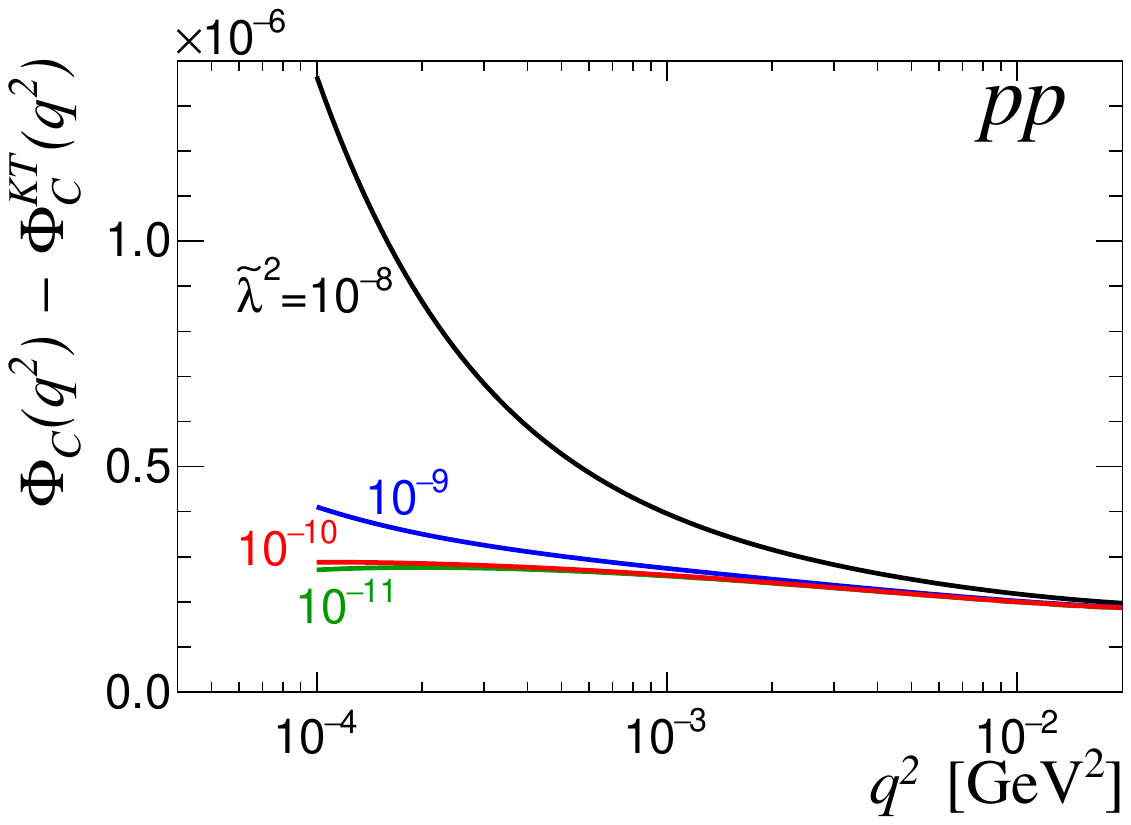}
  \end{center}
  \vspace{-2em}
  \caption{
    Comparison of the numerically computed Coulomb phase (dependent on photon mass $\lambda^2$) with the analytical result\,\cite{Kopeliovich:2000ez} obtained in the leading-order approximation.
    \label{fig:dPhiC}
  }
\end{figure}

Figure\,\ref{fig:dPhiC} compares the Coulomb phase $\Phi_C(q^2,\tilde{\lambda}^2)$ calculated numerically using Eq.\,\eqref{eq:FgC} with the analytical result from Ref.\,\cite{Kopeliovich:2000ez}. A small discrepancy of approximately $\sim2\times10^{-7}$ arises due to the leading-order $\alpha$ approximation in the analytical derivation. The photon-mass-dependent error in the numerical evaluation of $\Phi_C(q^2,\tilde{\lambda}^2)$ decreases rapidly with decreasing $\tilde{\lambda}^2$ and is on the order of $\mathcal{O}(10^{-8})$ for $\tilde{\lambda}^2\!=\!10^{-10}$.

The Coulomb correction to the hadronic amplitude~\eqref{eq:FgN} involves only a \textit{simple integration}, meaning that the integration range is naturally limited [see the $b$-dependence of $\gamma_N(b)$ in Fig.\,\ref{fig:fb}], and the integrand contains no singularities.

The resulting Coulomb phase $\Phi_N(q^2)$ is in approximate agreement with the analytical result $\Phi_N^\mathrm{KT}(q^2)$:
\begin{equation}
  \left| \Phi_N(q^2) - \Phi_N^\mathrm{KT}(q^2)\right| < 8\times10^{-9}.
\end{equation}
However, if the numerical calculation is also performed in the leading-order $\alpha$ approximation,
\begin{equation}
  \Phi_N^\text{LO}(q^2) = \frac%
      {\int db\,b\,\gamma_N(b)\,\chi'_C(b)\,J_0(qb)}%
      {\int db\,b\,\gamma_N(b)\,J_0(qb)},
\end{equation}
the result matches $\Phi_N^\mathrm{KT}(q^2)$ exactly, within computational precision ($\sim\!10^{-15}$).

\subsection{Coulomb correction in case of nonexponential form factor}

In the more general case of a nonexponential form factor $F_C(q^2)$, the form factor expression can be decomposed into exponential and nonexponential components:
\begin{equation}
  \frac{F_C(q^2)}{q^2} =
  \frac{e^{-B_Cq^2/2}}{q^2} +
  \frac{F_C(q^2)-e^{-B_Cq^2/2}}{q^2},
  \label{eq:FF}
\end{equation}
where
\begin{equation}
  B_C = \frac{-2\,d\ln{F_C(q^2)}}{dq^2}\Big|_{q^2=0}.
  \label{eq:B0}
\end{equation}
Since the nonexponential correction to the amplitude does not have a singularity at $q^2 \to 0$, the corresponding correction $\delta\chi_C(b)$ to the eikonal phase can be calculated without introducing a photon mass:
\begin{equation}
  \delta\chi_C(b) =\int_0^\infty{qdq\frac{F_C(q^2)-e^{-B_Cq^2/2}}{q^2} J_0(bq)}.
  \label{eq:dchiC}
\end{equation}

As an illustration, $\delta\chi_C(b)$ was evaluated for the frequently used dipole form of the proton electric form factor\,\cite{Chan:1966zza} $\left(1+q^2/\Lambda^2\right)^{-2}$, where $\Lambda^2 = 0.71\,\mathrm{GeV}^2$. In this case,
\begin{equation}
  F_C(q^2)=\left(1+q^2/\Lambda^2\right)^{-4},\quad\text{and}\quad B_C=8/\Lambda^2.
\end{equation}
The calculated $\delta\chi_C(b)$ is shown in Fig.\,\ref{fig:fb} as a dashed line. At $\tilde{b} \approx 2$, the function $\delta\chi_C(b)$ changes sign.

Thus, only {\em simple integration} is required to evaluate the effect of the nonexponential form factor in the Coulomb-corrected electromagnetic amplitude (in the massless photon approximation):
\begin{flalign}
  f_C^\gamma(q^2) &= \frac{e^{-B_Cq^2/2}}{q^2}\mathcal{F}_C(q^2)
  \nonumber \\ &+
  \int_0^\infty b\,db\,i\left[1-e^{i\delta\chi_C(b)}\right]e^{i\chi'_C(b)}J_0(bq).
  \label{eq:FgC_corr}
\end{flalign}
Here, $\mathcal{F}_C(q^2)$ and $\chi'_C(b)$ correspond to the exponential form factor with slope $B_C$, as defined in Eqs.\,\eqref{eq:FgC} and \eqref{eq:chiC'}, respectively.

If hadronic form factor $F_N(q^2)$ (with slope $B_N$ at $q^2\!\to\!0$), the Coulomb modified amplitude \eqref{eq:FgN} should be corrected as
\begin{flalign}
  &f_N^\gamma(q^2) = f_N(q^2)\mathcal{F}_N(q^2)
  \nonumber \\ &\qquad -
  \int_0^\infty b\,db\,\gamma_N(b)\left[1-e^{i\delta\chi_C(b)}\right]e^{i\chi'_C(b)}J_0(bq).
  \nonumber \\ &\qquad +
  \int_0^\infty b\,db\,\delta\gamma_N(b)e^{i\delta\chi_C(b)}\,e^{i\chi'_C(b)}J_0(bq),
  \label{eq:FgN_corr}
\end{flalign}
where 
\begin{equation}
  \delta\gamma_N(b) =\int_0^\infty{qdq \left[F_N(q^2)-e^{-B_Nq^2/2}\right] J_0(bq)}.
  \label{eq:dchiN}
\end{equation}

\section{Spin-flip proton-proton scattering \label{sec:pp_sf} }

In general, the single spin-flip amplitude for transversely polarized proton scattering contains azimuthal angle factors: $\cos{\varphi_q} = (\boldsymbol{n\cdot{q}})/q$ in momentum space and $\cos{\varphi_b} = (\boldsymbol{n\cdot{b}})/b$ in impact parameter space. As shown in Appendix\,\ref{sec:Fourier}, these factors transform into each other under Fourier transformation:
\begin{equation}
  \cos{\varphi_q} \to i\cos{\varphi_b} \to \cos{\varphi_q}.
\end{equation}
For simplicity, we omit these angular factors in the expressions for the amplitudes. In particular, the hadronic spin-flip amplitude $f_S(q^2)$ and the electromagnetic spin-flip amplitude $f_M(q^2)$ are written as
\begin{align}
  f_S(q^2) &= \frac{\sigma_\mathrm{tot} r_5q}{4\pi m_p} e^{-B_Sq^2/2}, \\
  f_M(q^2) &= \frac{-\kappa_p \alpha}{m_p q} e^{-B_M q^2/2}.
\end{align}
However, it should be kept in mind that an additional angular factor may be required in cases involving multiple spin-flip amplitudes, as demonstrated in Eqs.\,\eqref{eq:cos2} and \eqref{eq:cos3}.

Considering $p^{\uparrow}p$ scattering in the eikonal model, both the electromagnetic $\chi_M(b)$ and hadronic $\chi_S(b)$ spin-flip phases must be included in the profile function:
\begin{flalign}
  i\Gamma &= i\left[1 - e^{i\chi_N(b) + i\chi_S(b) + i\chi_C(b) + i\chi_M(b)}\right]
  \nonumber \\ &\to
  i\left[1 - e^{i\chi_C(b)}\right]e^{-i\Delta}
  \nonumber \\ &+
  \left[\gamma_N + \gamma_S + \gamma_M + i\gamma_N \gamma_M + i\gamma_S \gamma_M\right]e^{i\chi'_C(b)},
  \label{eq:Gsf}
\end{flalign}
where the nonflip components $\chi_C(b)$ and $\gamma_N(b)$ were defined earlier. Based on the results derived in Appendix\,\ref{sec:Fourier}, the corresponding spin-flip components take the form
\begin{flalign}
  \gamma_S(b) &= i\left[1 - e^{i\chi_N(b) + i\chi_S(b)}\right] - \gamma_N(b)
  \nonumber \\ &=
  \frac{\sigma_\mathrm{tot} r_5}{4\pi m_p} \frac{b e^{-b^2/2B_S}}{B_S^2}, \\
  \gamma_M(b) &= i\left[1 - e^{i\chi_M(b)}\right] 
  \nonumber \\ &\approx
  \chi_M(b) = \frac{-\alpha \kappa_p}{m_p} \frac{1 - e^{-b^2/2B_M}}{b}.
  \label{eq:chiM}
\end{flalign}

The term $i\gamma_S(b)\gamma_M(b)$ effectively alters the hadronic nonflip amplitude. However, due to its smallness --- of order $\sim \alpha r_5 q^2/m_p^2$ relative to $\gamma_N$ --- it can be neglected.

The single spin-flip term $i\gamma_N(b)\gamma_M(b)$, which can be interpreted either as a magnetic photon correction to the hadronic nonflip amplitude or as an absorptive correction to the electromagnetic spin-flip amplitude, can become significant in proton-nucleus scattering involving large nuclear charge $Z$, such as in $p^\uparrow\mathrm{Au}$ scattering. A more detailed discussion of this term is provided in Sec.\,\ref{sec:abs}.

The Coulomb correction to the hadronic spin-flip amplitude,
\begin{align}
  f_S^\gamma(q^2) &= \int_0^\infty b\,db\,\gamma_S(b) e^{i\chi'_C(b)} J_1(bq)
  \nonumber \\
  &= f_S(q^2) \times \mathcal{F}_S(q^2),
  \label{eq:FgS}
\end{align}
meets the criterion of {\em simple integration} and can thus be calculated straightforwardly. In case of nonexponential form factor $F_S(q^2)$, $\gamma_S(b)$ is altered by
\begin{equation}
  \delta\gamma_S(b) =\int_0^\infty{q^2dq \left[F_S(q^2)-e^{-B_Sq^2/2}\right] J_1(bq)}
  \label{eq:dchiS}
\end{equation}
and 
\begin{flalign}
  &f_S^\gamma(q^2) = f_S(q^2)\mathcal{F}_S(q^2)
  \nonumber \\ &\qquad -
  \int_0^\infty b\,db\,\gamma_S(b)\left[1-e^{i\delta\chi_C(b)}\right]e^{i\chi'_C(b)}J_1(bq).
  \nonumber \\ &\qquad +
  \int_0^\infty b\,db\,\delta\gamma_S(b)e^{i\delta\chi_C(b)}\,e^{i\chi'_C(b)}J_1(bq),
  \label{eq:FgS_corr}
\end{flalign}

A potential difficulty in calculating the Coulomb correction to the electromagnetic spin-flip amplitude,
\begin{align}
  f_M^\gamma(q^2) &= \int_0^\infty bdb\,\chi_M(b) e^{i\chi'_C(b)} J_1(bq)
  \nonumber \\
  &= f_M(q^2) \times \mathcal{F}_M(q^2),
  \label{eq:FgM}
\end{align}
is the slow, $\sim 1/b$, decrease of $\chi_M(b)$ at large $b$, which requires numerical integration over a wide range of $b$. Nevertheless, a compact expression allowing precise calculation of the Coulomb correction to the magnetic amplitude was derived in Appendix\,\ref{sec:FgM}:
\begin{equation}
  \mathcal{F}_M(q^2) = \widetilde{\mathcal{F}}(B_C q^2/2,B_C/B_M).
  \label{eq:FgM_}
\end{equation}

\begin{figure}[t]
  \begin{center}
    \includegraphics[width=0.85\columnwidth]{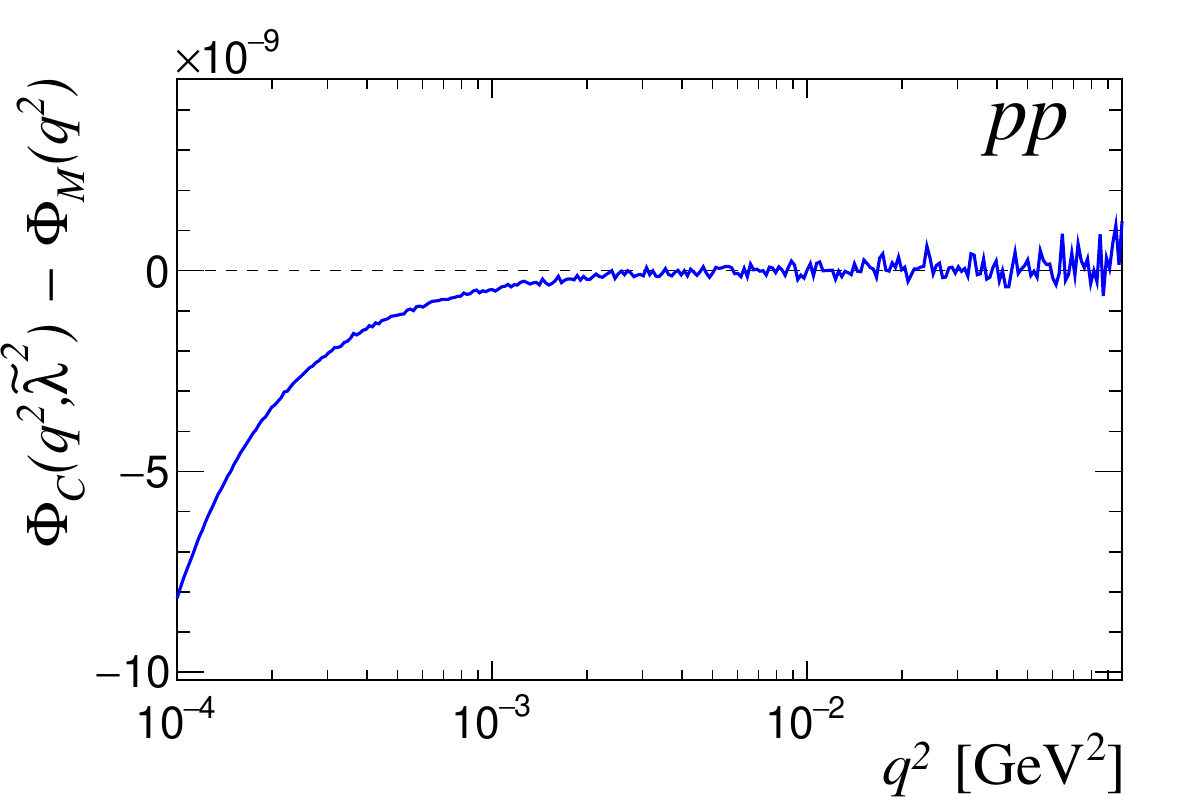}
    \includegraphics[width=0.85\columnwidth]{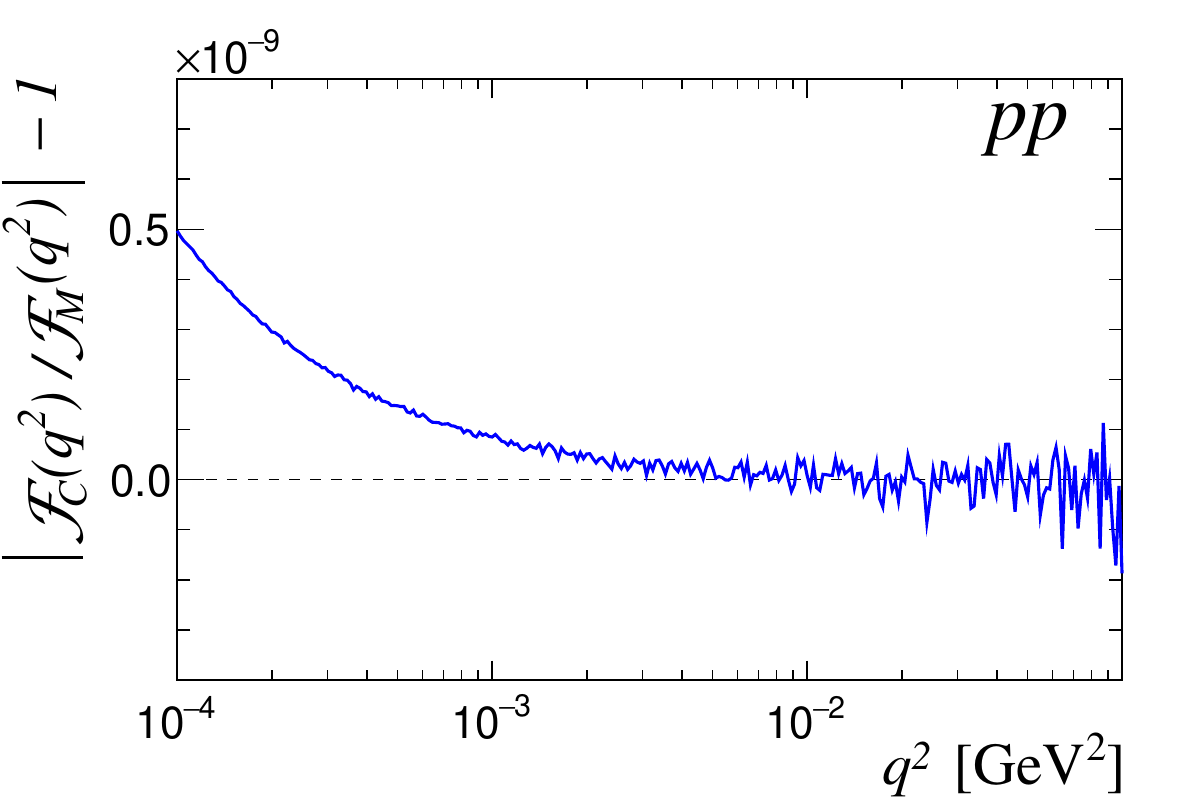}
  \end{center}
  \vspace{-2em}
  \caption{
    \label{fig:C-M_1}
    Comparison of the nonflip $|\mathcal{F}(q^2)|e^{i\Phi_C(q^2)}$ and spin-flip $|\mathcal{F}(q^2)|e^{i\Phi_C(q^2)}$ Coulomb corrections to the forward elastic $p^{\uparrow}p$ electromagnetic amplitudes.
  }
\end{figure}

For the nonexponential electromagnetic spin-flip form factor $F_M(q^2)$, 
\begin{equation}
  \delta\gamma_M(b) =\int_0^\infty{dq \left[F_M(q^2)-e^{-B_Mq^2/2}\right] J_1(bq)}
  \label{eq:dchiM}
\end{equation}
and 
\begin{flalign}
  &f_M^\gamma(q^2) = f_M(q^2)\mathcal{F}_M(q^2)
  \nonumber \\ &\qquad -
  \int_0^\infty bdb\,\gamma_M(b)\left[1-e^{i\delta\chi_C(b)}\right]e^{i\chi'_C(b)}J_1(bq).
  \nonumber \\ &\qquad +
  \int_0^\infty bdb\,\delta\gamma_M(b)e^{i\delta\chi_C(b)}\,e^{i\chi'_C(b)}J_1(bq).
  \label{eq:FgM_corr}
\end{flalign}

In Ref.\,\cite{Buttimore:1978ry}, it was found that the Coulomb phase is insensitive to the spin of the particles involved and is therefore identical across all helicity amplitudes. In particular, one may expect that
\begin{equation}
  \mathcal{F}_M(q^2) = \mathcal{F}_C(q^2).
  \label{eq:FgM=FgC}
\end{equation}

Using the analytical calculation\,\cite{Kopeliovich:2000ez} of the Coulomb phase \(\Phi_C^\text{LO}(q^2)\) in the leading-order approximation, one can readily show\,\cite{Poblaguev:2021xkd} that \(\Phi_M^\text{LO}(q^2) = \Phi_C^\text{LO}(q^2)\). Furthermore, our numerical calculation of the spin-flip phase \(\Phi_M^\text{LO}(q^2)\), given by Eq.\,\eqref{eq:PhiM_LO}, exactly matches the nonflip phase \(\Phi_{C}^\text{KT}(q^2)\).

For all orders in $\alpha$, the numerically computed Coulomb corrections for the spin-flip and nonflip amplitudes are compared in Fig.\,\ref{fig:C-M_1}. Within the momentum transfer squared range relevant to HJET measurements, the consistency is better than $10^{-9}$. A more conservative conclusion is that $\mathcal{F}_M(q^2)$ and $\mathcal{F}_C(q^2)$ are identical within the CNI region to a precision of $10^{-8}$. Although a small $q^2$-dependent discrepancy is observed between the nonflip and spin-flip Coulomb corrections, it has not been conclusively shown that this difference is not due to systematic errors in the numerical calculation of $\mathcal{F}_C(q^2)$. Moreover, beyond the CNI region, one of the amplitudes --- electromagnetic or hadronic --- is effectively suppressed, and as a result, the significance of the Coulomb correction is substantially reduced.

Thus, given the actual experimental accuracy of the HJET measurements, the equality expressed in Eq.\,\eqref{eq:FgM=FgC} is well-justified.

\section{Proton-Nucleus Scattering}

\begin{figure*}[t]
  \begin{center}
    \includegraphics[width=0.325\textwidth]{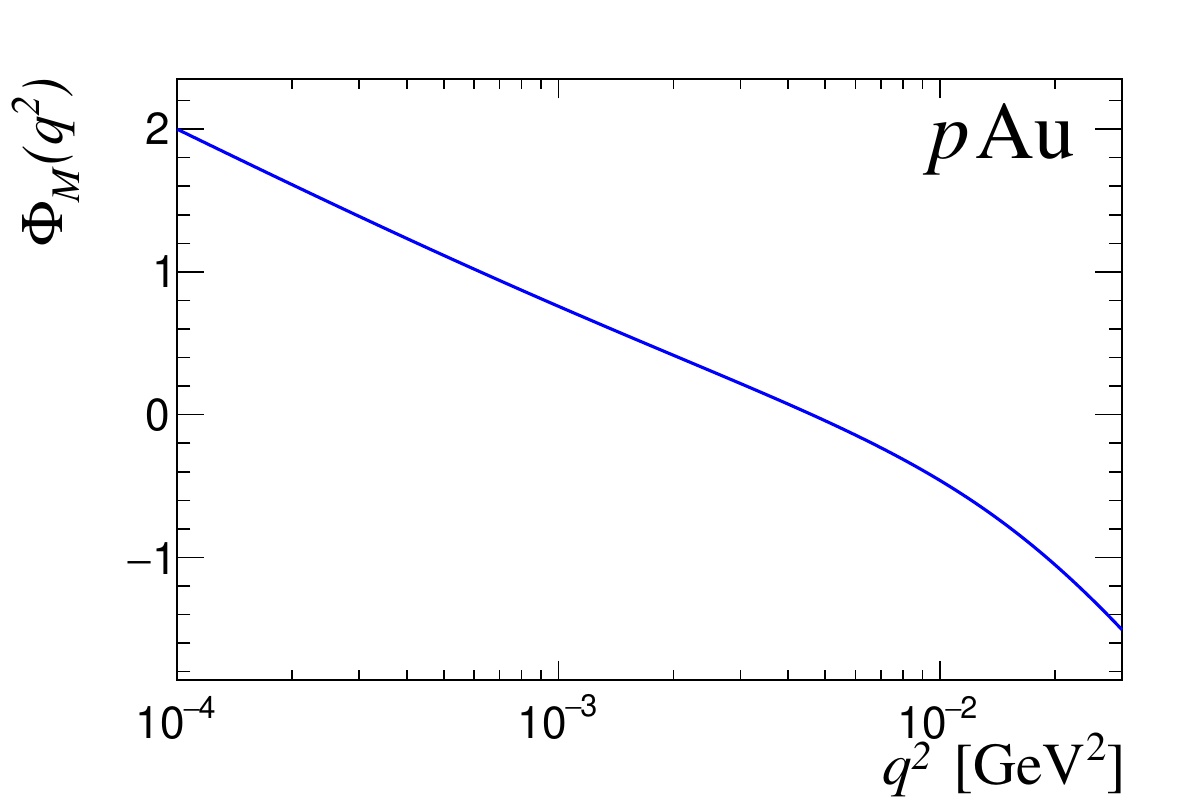} \hfill
    \includegraphics[width=0.325\textwidth]{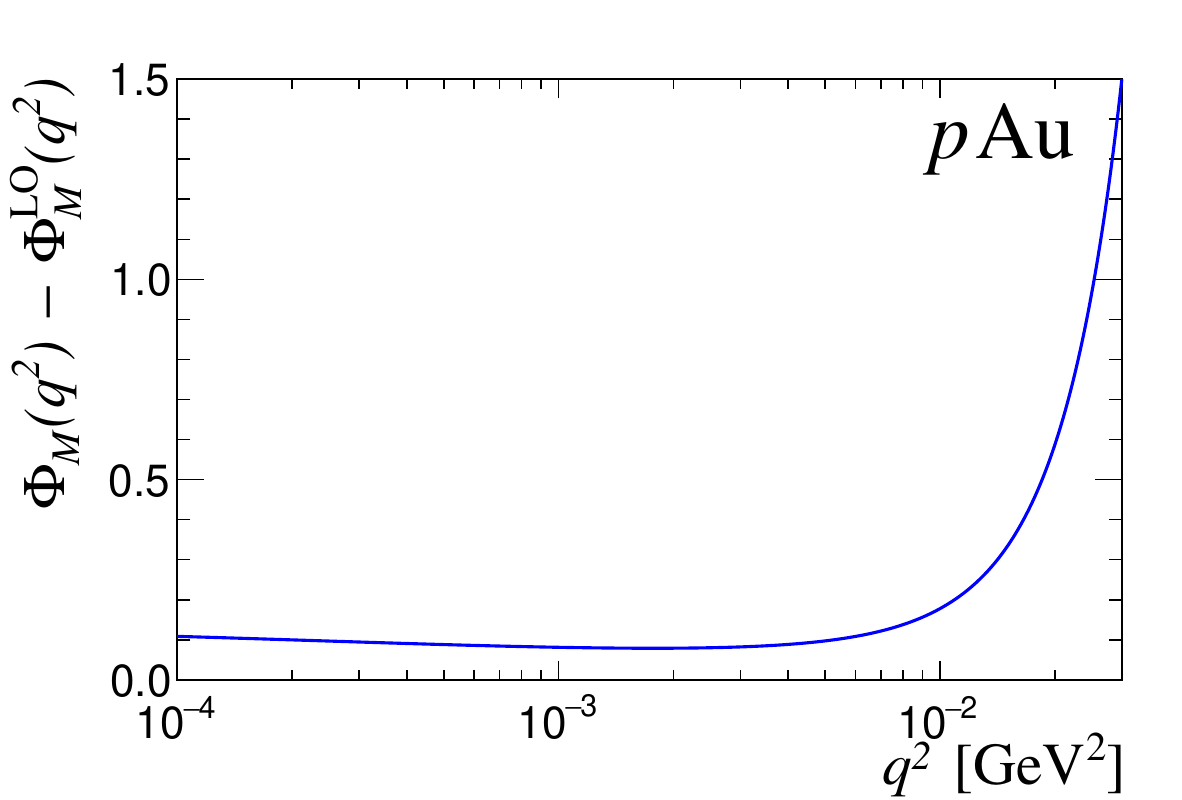} \hfill
    \includegraphics[width=0.325\textwidth]{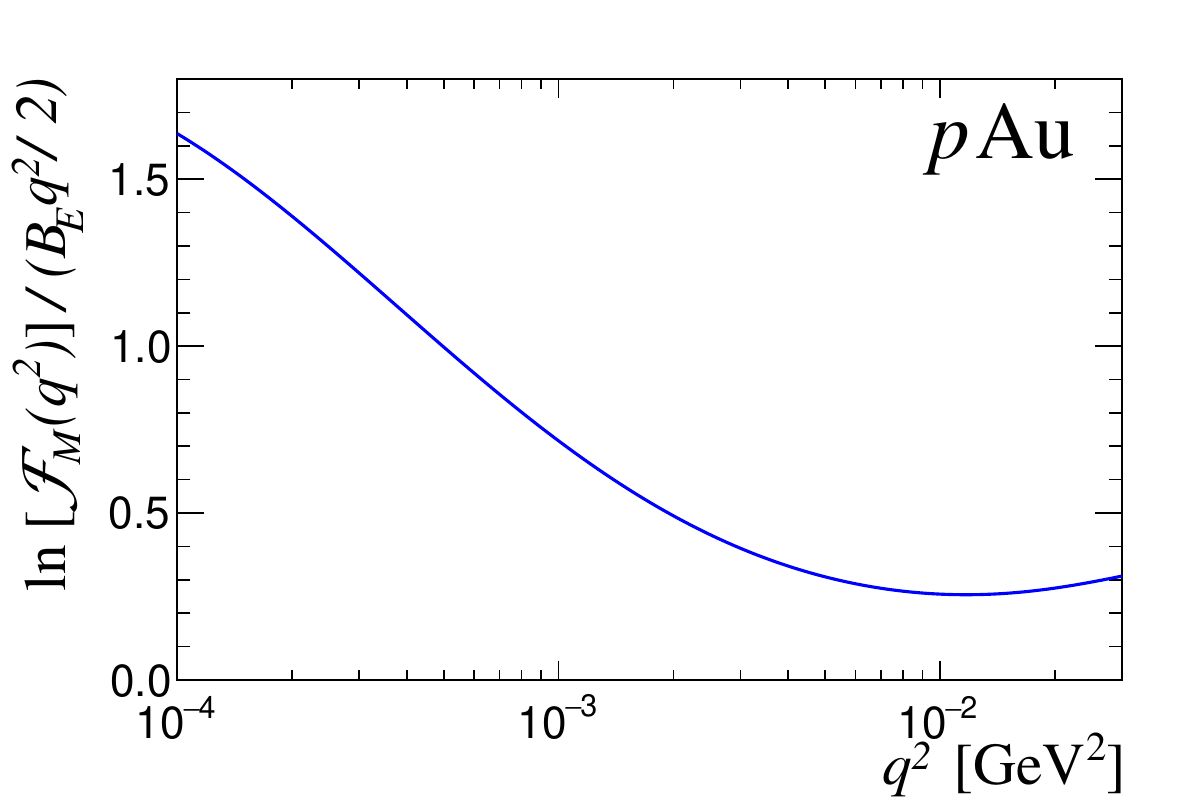} 
  \end{center}
  \vspace{-2em}
  \caption{
    Evaluation of the Coulomb correction, $\mathcal{F}_M(q^2)=\left|\mathcal{F}_M(q^2)\right|\times\exp{(i\Phi_M(q^2))}$, to the $p^\uparrow$Au electromagnetic spin-flip amplitude.
    \label{fig:pA_Z=79}
  }
\end{figure*}

\begin{figure}[t]
  \begin{center}
    \includegraphics[width=0.85\columnwidth]{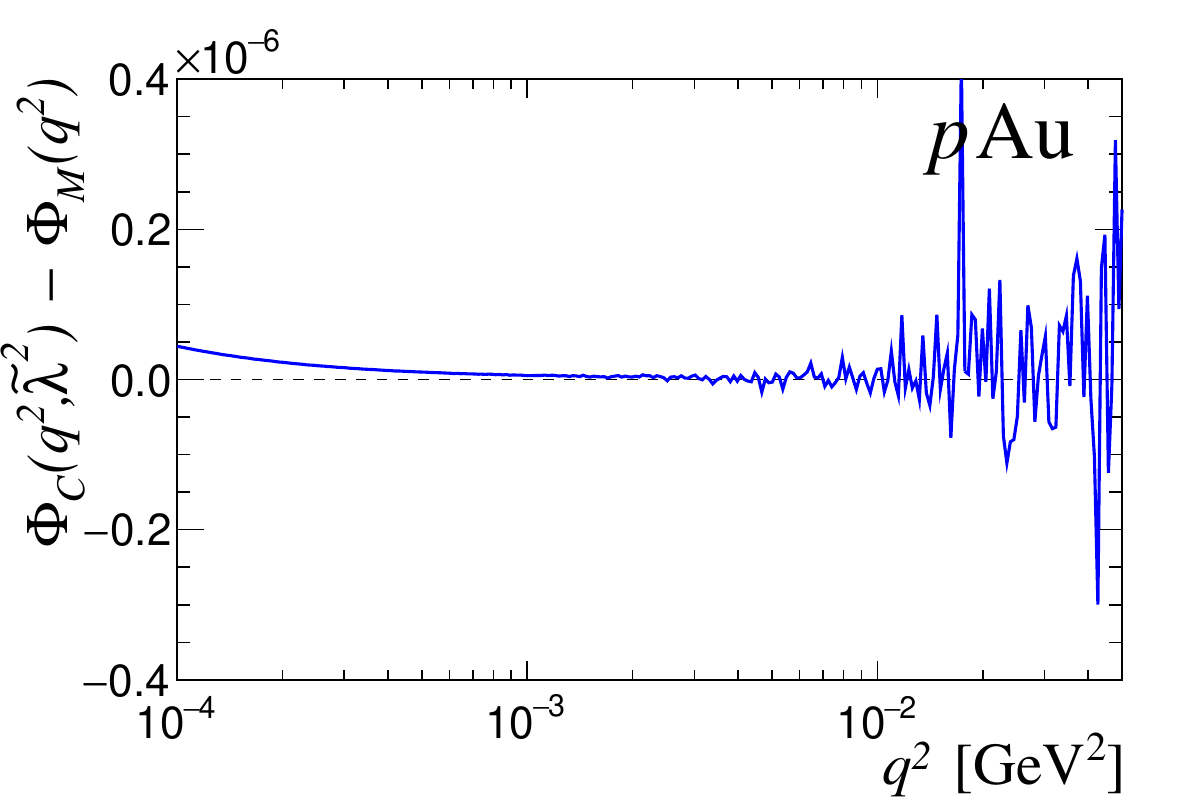}
    \includegraphics[width=0.85\columnwidth]{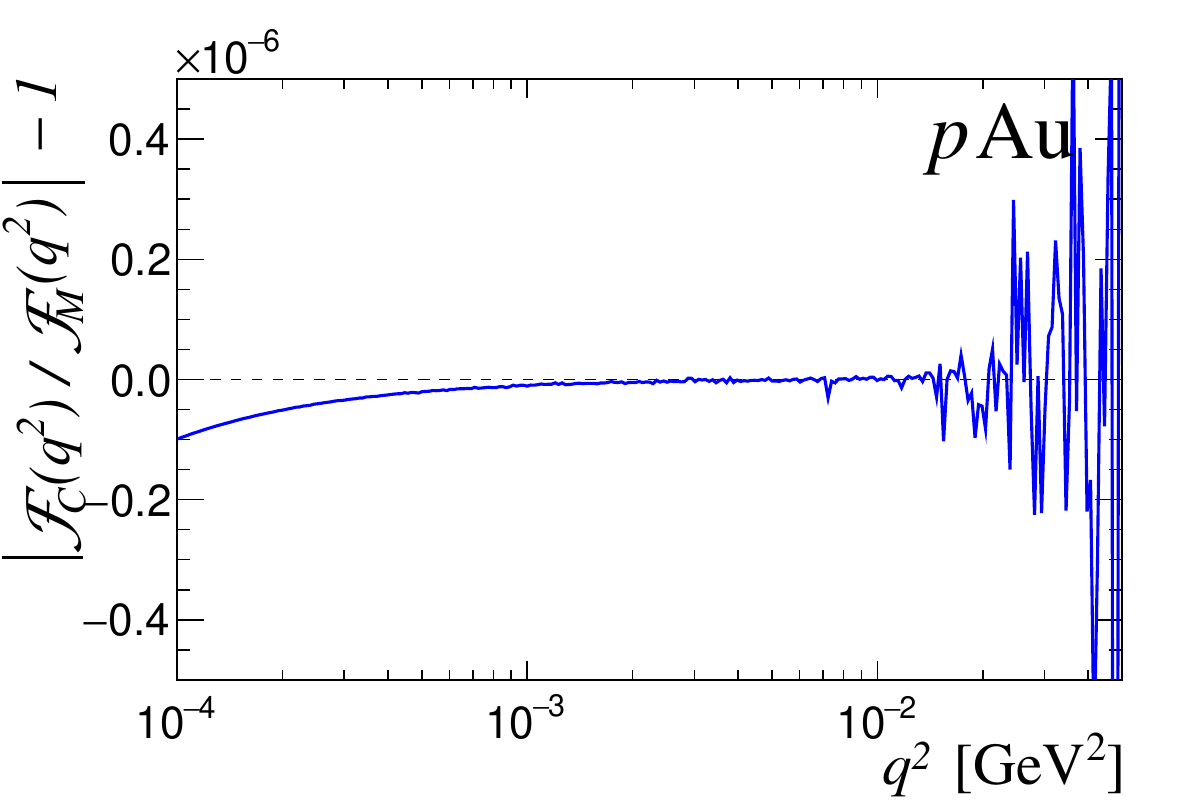}
  \end{center}
  \vspace{-2em}
  \caption{
    \label{fig:C-M_79}
    Comparison of the Coulomb corrections to the nonflip and spin-flip forward elastic $p^\uparrow$Au electromagnetic amplitudes: $|\mathcal{F}_C(q^2)|e^{i\Phi_C(q^2)}$ vs.\ $|\mathcal{F}_M(q^2)|e^{i\Phi_M(q^2)}$. The instability observed at large $q^2$ is attributed to numerical calculation uncertainties in the evaluation of $\mathcal{F}_C(q^2)$.
  }
\end{figure}

The equations used in Secs.\,\ref{sec:pp_nf} and \ref{sec:pp_sf} to evaluate the Coulomb correction in proton-proton scattering are also applicable to proton-nucleus scattering, provided that $\alpha$ is replaced by $\alpha Z$ and the experimentally determined parameters (e.g., $\rho$, $\sigma_\text{tot}$, and $B_X$) are adjusted accordingly. In Ref.\,\cite{Kopeliovich:2023xtu}. it was suggested to use Glauber model to evaluate $\rho^{pA}$, $\sigma_\text{tot}^{pA}$, and $F_X^{pA}(q^2)$.

Here, to estimate the Coulomb corrections to the electromagnetic $p$Au amplitude ($Z = 79$), an exponential form factor with $B_C = 250\,\text{GeV}^{-2}$ was assumed. It should be emphasized that this is an oversimplified approach (e.g., diffraction effects are not included), and the resulting estimates primarily serve to illustrate the scale of Coulomb corrections in $p$Au scattering.

Estimates for the spin-flip amplitude are shown in Fig.\,\ref{fig:pA_Z=79}. Notably: {\em(i)} the Coulomb phase increases substantially due to the large nuclear charge $Z$, and it varies significantly over the HJET momentum transfer range; {\em(ii)} higher-order corrections in $\alpha$ are important for accurately calculating $\Phi_M(q^2)$; {\em(iii)} the effective correction to the form factor slope, given by $\ln{[\mathcal{F}_M(q^2)]}/(B_M q^2/2)$, is comparable in magnitude to the slope itself.

A comparison of the nonflip and spin-flip Coulomb corrections for $p$Au scattering is presented in Fig.\,\ref{fig:C-M_79}. While the difference between $\mathcal{F}_C(q^2)$ and $\mathcal{F}_M(q^2)$ in the CNI region is larger, on the order of $\sim 10^{-7}$, than that observed in $p^\uparrow p$ scattering, it remains negligible for practical purposes in experimental data analysis.

For all nuclei shown in Fig.\,\ref{fig:pA}, a more detailed evaluation of the Coulomb corrections --- including Glauber-based calculations of the $pA$ form factors and the effects of soft magnetic photon exchanges (absorptive corrections) --- will be presented in an upcoming publication by the RHIC Polarimetry Group on $pA$ analyzing power measurements at HJET.

\section{The absorptive correction  \label{sec:abs} }

Following Ref.\,\cite{Kopeliovich:2023xtu}, the term $i\gamma_N(b)\gamma_M(b)$ in Eq.\,\eqref{eq:Gsf} is referred to as the absorptive correction. In the case of exponential form factors, this correction can be expressed as the difference between two spin-flip electromagnetic amplitudes:
\begin{align}
  \gamma_\text{NM}(b) &= i\gamma_N(b)\gamma_M(b) \nonumber \\
   &= C_\text{NM} \times%
  \left[\hat{\gamma}_M(b, B_\text{NM}) - \hat{\gamma}_M(b, B_N)\right],
\end{align}
where
\begin{align}
 C_\text{NM} &= (1 - i\rho)\frac{\sigma_\text{tot}}{4\pi B_N}\frac{\alpha Z \kappa_p}{m_p}, \\
  B_\text{NM} &= \frac{B_N B_M}{B_N + B_M}.
\end{align}

The corresponding effective amplitude in momentum space, including Coulomb corrections, is given by
\begin{flalign}
  f_\text{NM}^\gamma(q) = C_\text{NM}\,q^{-1}\times&%
  \Big[ e^{-B_\text{NM}q^2/2} \widetilde{\mathcal{F}}\left(B_Cq^2/2,B_C/B_\text{NM}\right)
    \nonumber \\&-
    e^{-B_Nq^2/2} \widetilde{\mathcal{F}}\left(B_Cq^2/2,B_C/B_N\right)
\Big]\!.
 \label{eq:fNM}
\end{flalign}

In case of nonexponential form factors $F_N(q^2)$ and $F_M(q^2)$, 
\begin{equation}
  \delta\gamma_\text{NM}(b) = C_\text{NM}\times\left[%
      \gamma_N \delta\gamma_M + \delta\gamma_N\gamma_M +%
      \delta\gamma_N \delta\gamma_M \right],
\end{equation}
where $\delta\gamma_N(b)$ and $\delta\gamma_M(b)$ are defined in Eqs.\,\eqref{eq:dchiN} and \eqref{eq:dchiM}. Consequently, nonexponential corrections to  $f_\text{NM}^\gamma(q^2)$ can be calculated by substitution of $\gamma_\text{NM}(b) $ and $\delta\gamma_\text{NM}(b)$ to Eq.\,\eqref{eq:FgM_corr} 

To evaluate the significance of the absorptive correction, a simplified case with $B_N = B_M$ and $\widetilde{\mathcal{F}}(q^2) = 1$ can be considered. In this scenario, the correction can be interpreted either as an effective modification of the proton's anomalous magnetic moment,
\begin{equation}
  \kappa_p \to \kappa_p^\text{eff} = \kappa_p \times%
  \left(1 - \frac{\alpha Z}{2} \frac{t}{t_c}\right),
  \label{eq:kpZ}
\end{equation}
or as a shift in the hadronic spin-flip amplitude:
\begin{equation}
  r_5 \to r_5^\text{eff} = r_5 + \frac{\alpha Z \kappa_p}{4}.
\end{equation}

For proton-proton scattering ($Z = 1$), the correction is small but not negligible\,\cite{Poblaguev:2021xkd}:
\begin{equation}
  \mathrm{Re}\,r_5^{pp} \to \mathrm{Re}\,r_5^{pp} + 0.003.
\end{equation}
However, in proton-gold scattering ($Z = 79$), the absorptive correction has a significant impact on the analyzing power $A_\text{N}^{p\mathrm{Au}}(t)$. In particular, absorption explains the rapid decrease of $A_\text{N}^{p\mathrm{Au}}(t)$ at small $t$, as shown in Fig.\,\ref{fig:pA}.

  The zero of $A_\text{N}^{p\mathrm{Au}}(t)$ at $|t| \!\approx\! 0.012\,\text{GeV}^2$ can be attributed --- consistent with Eq.\,\eqref{eq:AN} --- to the vanishing of $\mathrm{Im}\,F_{nf}$ due to diffraction. In contrast, the sign change of the analyzing power at $|t| \!\approx\! 0.005\,\text{GeV}^2$ results from the condition $F_{sf} = 0$, reflecting the effective suppression of the electromagnetic spin-flip amplitude caused by soft magnetic photon exchange (i.e., absorption). As shown in Fig.\,\ref{fig:pA} and in agreement with Eq.\,\eqref{eq:kpZ}, the strength of this absorptive correction exhibits a strong dependence on the nuclear charge $Z$. It should be also noted that significance of the absorptive corrections for interpretation of the forward polarized $p^\uparrow$Au scattering was underlined in Refs.\,\cite{Krelina:2019mlu,Kopeliovich:2023xtu}.

\section{Multiple Magnetic Photon Exchange  \label{sec:MF}}

\begin{figure}[t]
  \begin{center}
    \includegraphics[width=0.85\columnwidth]{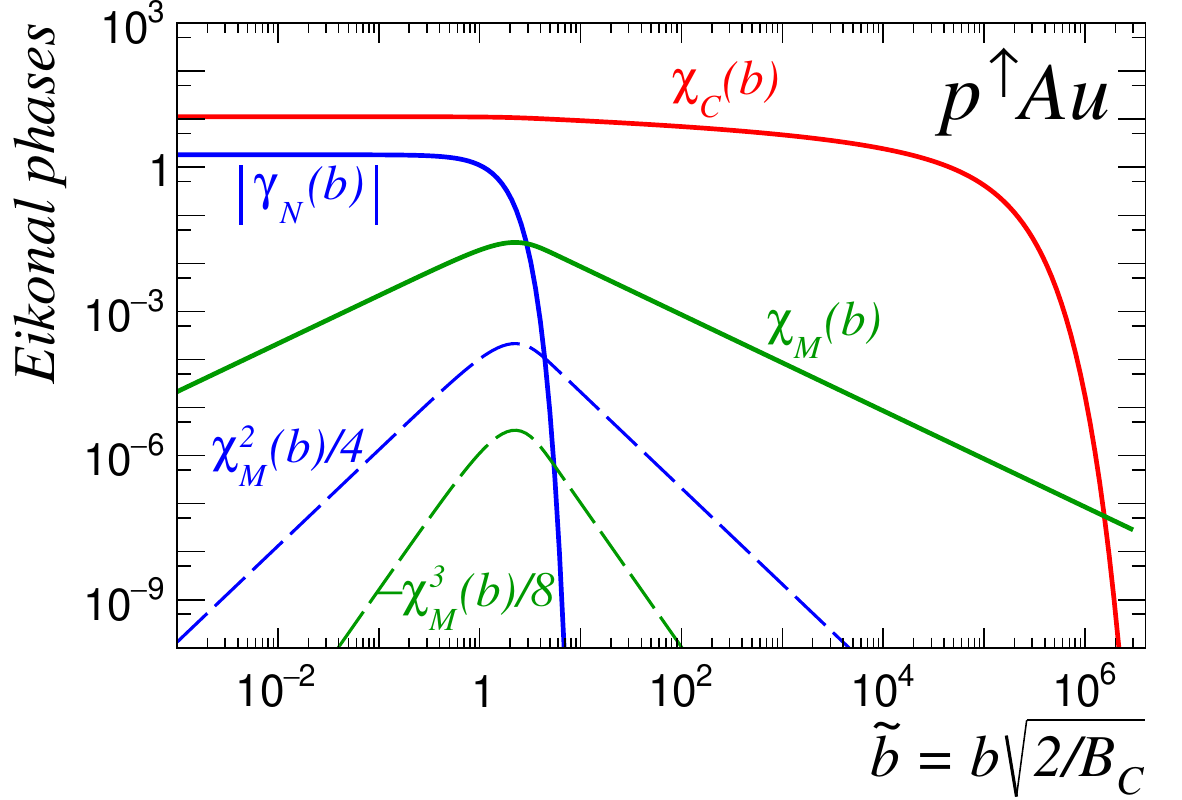}
  \end{center}
  \vspace{-2em}
  \caption{
    Eikonal phases in $p^\uparrow$Au scattering as functions of the impact parameter $b$.  
    \label{fig:fbAu}
  }
\end{figure}

\begin{figure}[t]
  \begin{center}
    \includegraphics[width=0.85\columnwidth]{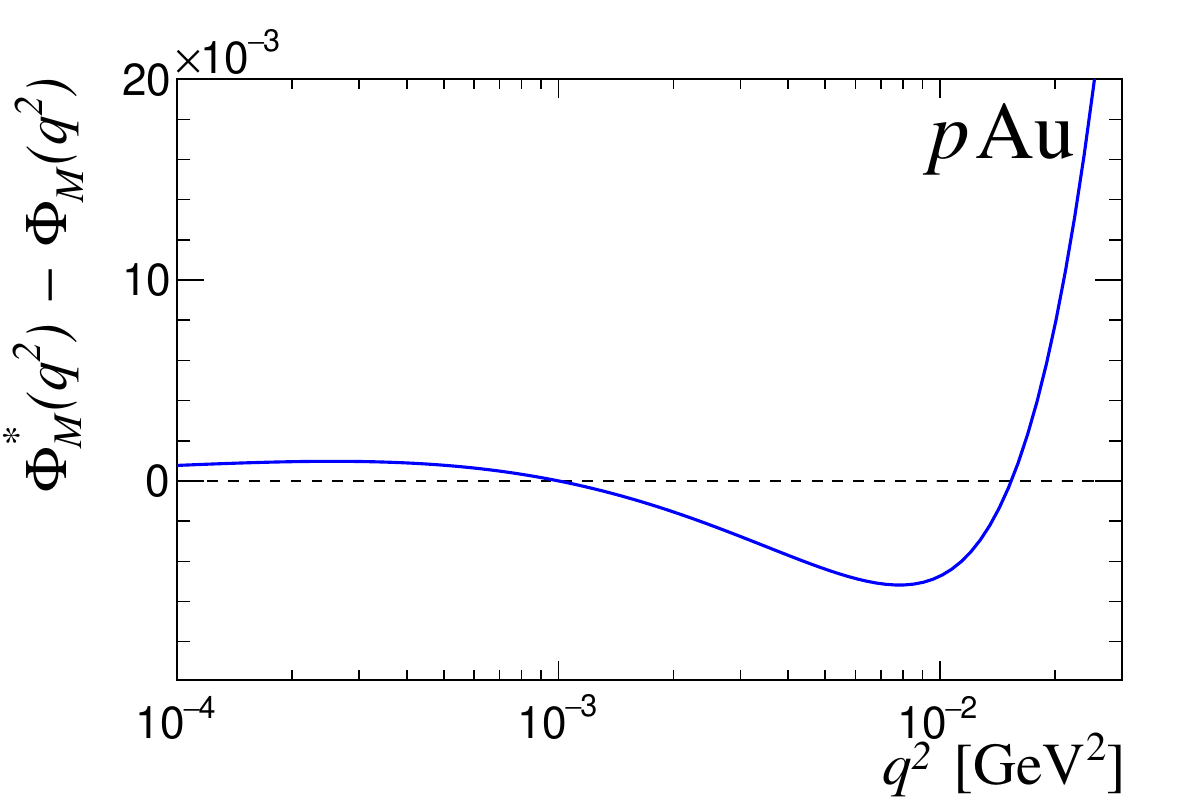}
    \includegraphics[width=0.85\columnwidth]{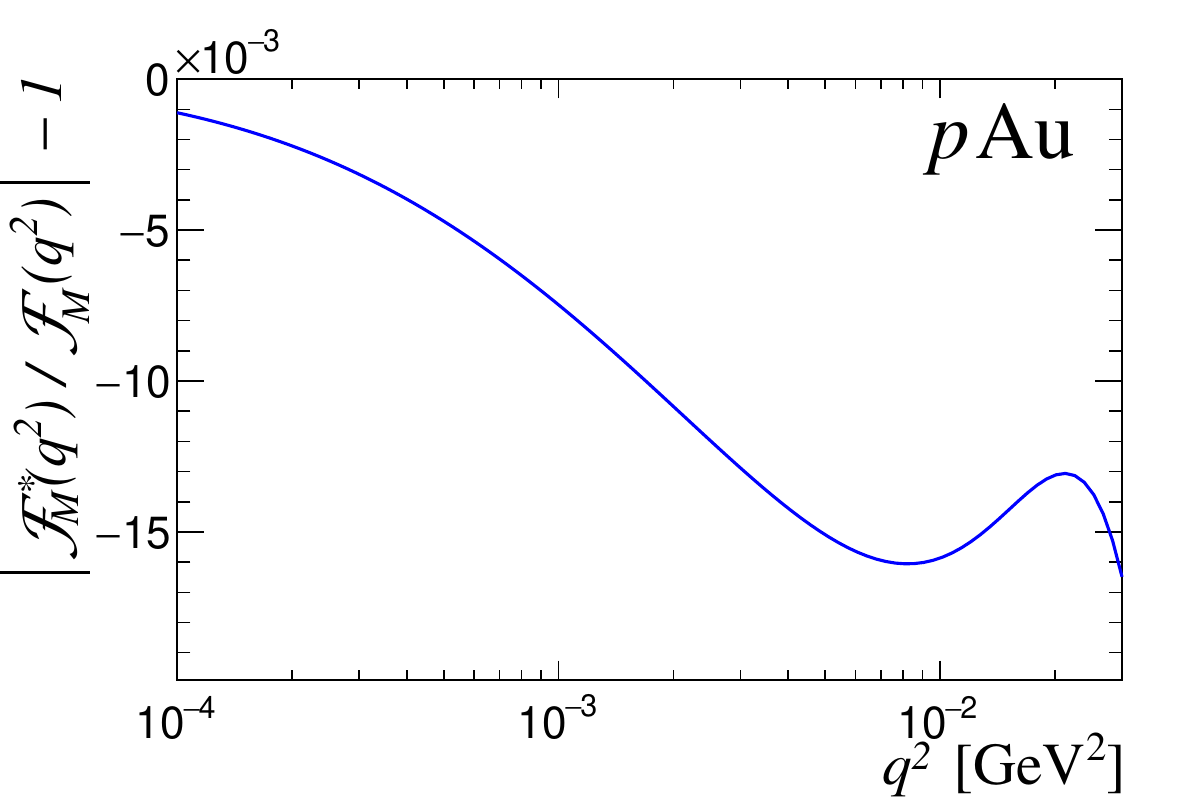}
  \end{center}
  \vspace{-2em}
  \caption{
   Comparison of the combined correction $\mathcal{F}^*_M(q^2)$ --- which includes both Coulomb and higher-order $\chi_M(b)$ effects --- to the $p^\uparrow$Au electromagnetic spin-flip amplitude, with the Coulomb-only correction $\mathcal{F}_M(q^2)$ [Eq.\,\eqref{eq:FgM}].
  \label{fig:M3}
}
\end{figure}

Previously, in Eq.\,\eqref{eq:chiM}, the electromagnetic spin-flip eikonal amplitude $\gamma_M(b)$ was approximated by the eikonal phase $\chi_M(b)$, corresponding to the leading term in its expansion. Including higher-order terms and accounting for angular factors from Eqs.\,\eqref{eq:cos2} and \eqref{eq:cos3}, both spin-flip and nonflip components arise:
\begin{equation}
  \gamma_M(b) = \chi_M\left(1 - \frac{\chi_M^2}{8}\right)\Big|_\text{spin-flip}
  + i\frac{\chi_M^2}{4}\Big|_\text{nonflip}.
\end{equation}

In forward $pp$ scattering ($Z\!=\!1$), the correction to the leading-order approximation is of order $(\alpha Z)^2 q^2 / m_p^2$ and can thus be safely neglected.

For $p^\uparrow$Au scattering, the relative contributions of the eikonal amplitude components are shown in Fig.\,\ref{fig:fbAu}. The specific contribution of the $\chi_M^3(b)/8$ term to the total correction of the electromagnetic spin-flip amplitude is evaluated in Fig.\,\ref{fig:M3}. Although the effect is small, it may warrant inclusion depending on the required precision of experimental measurements.

\section{Summary}
  
The possibility of numerically calculating Coulomb corrections to the nonflip and spin-flip amplitudes in forward elastic scattering of transversely polarized protons off proton or nuclear targets was investigated within the eikonal framework. It was assumed that all relevant form factors --- electromagnetic nonflip $F_C(t)$ and spin-flip $F_M(t)$, and hadronic nonflip $F_N(t)$ and spin-flip $F_S(t)$ --- are known in the Born approximation.

For computational convenience, it is useful to separate the exponential and nonexponential components of the form factors, as done in Eqs.\,\eqref{eq:FF} and \eqref{eq:B0}.

For exponential form factors, analytical expressions for the corresponding eikonal phases were derived. Inverse Fourier transformation --- including convolution with $\exp[i\chi_C(b,\lambda^2)]$, the term describing multiple soft photon exchange --- yields the Coulomb-corrected amplitudes for $p^{\uparrow}p$ and $p^{\uparrow}A$ scattering.

To avoid infrared divergences, the Coulomb phase $\chi_C(b,\lambda^2)$ was calculated using a fictitious photon mass $\lambda$. Reducing $\lambda$ improves the accuracy of the Coulomb correction $\mathcal{F}_C(q^2,\lambda^2)$; however, very small values of $\lambda$ require large integration ranges, which increases computation time and may also introduce numerical errors due to the limitations of floating-point arithmetic (even in double precision).

In contrast, the Coulomb correction $\mathcal{F}_M(q^2)$ [Eq.\,\eqref{eq:FgM_}] to the spin-flip electromagnetic amplitude can be computed directly in the massless limit $\lambda = 0$. In this case, $\chi_C(b,\lambda^2)$ is replaced by the compact, modified phase $\chi'_C(b)$ [Eq.\,\eqref{eq:chiC'}], and the numerical evaluation of $\mathcal{F}_M(q^2)$ reduces to a {\em simple integration} \eqref{eq:FM}.

Coulomb corrections to the nonflip and spin-flip electromagnetic amplitudes are expected\,\cite{Buttimore:1978ry} to be identical, which was confirmed by numerical calculations with accuracy better than $10^{-7}$. Thus, $\mathcal{F}_C(q^2,\lambda^2\!\to\!0)$ can be well-approximated as
\begin{equation}
  \mathcal{F}_C(q^2) = \widetilde{\mathcal{F}}(B_Cq^2/2,1),
\end{equation}
calculated using the exponential nonflip form factor.

This approach allows for accurate calculation of Coulomb corrections to all exponential amplitudes in the $\lambda \to 0$ limit, i.e., using the modified Coulomb phase $\chi'_C(b)$. Corrections due to nonexponential components can also be efficiently incorporated, as demonstrated in Eqs.\,\eqref{eq:FgC_corr}, \eqref{eq:FgN_corr}, \eqref{eq:FgS_corr}, and \eqref{eq:FgM_corr}.

In addition to Coulomb corrections, the scattering amplitude may also be influenced by soft magnetic photon exchange. This introduces a correction to the hadronic nonflip amplitude, generating an effective spin-flip amplitude $f_\text{NM}(q^2)$. For large $Z$, this contribution may exceed the ``standard'' spin-flip amplitudes $f_M(q^2)$ and $f_S(q^2)$, and must therefore be carefully accounted for in data analysis. As shown in Sec.\,\ref{sec:abs}, $f_\text{NM}(q^2)$ can be approximated by the difference of two electromagnetic spin-flip amplitudes, making it amenable to precise calculation --- including Coulomb corrections --- within the eikonal model.

For heavy nuclei (large $Z$), multiple soft magnetic photon exchanges may also need to be considered, depending on the required precision of analyzing power measurements. The necessity of such corrections was evaluated in Sec.\,\ref{sec:MF}.

\appendix

\section{Mathematical Relations Used \label{sec:Integrals}}

A typical integral used in the Fourier transformation within the eikonal approach is  
\begin{equation}
  f(\boldsymbol{y}) =%
  \frac{1}{2\pi} \int d^2x\,e^{\pm i\,\boldsymbol{x}\cdot\boldsymbol{y}}%
    (\boldsymbol{n}\cdot\boldsymbol{x})^k x^p e^{-a x^2},
  \label{eq:FourierIntegral}
\end{equation}
where $k \!\geq\! 0$ and $p$ are integers, and $\boldsymbol{n}$ is a unit vector in the $xy$-plane.

Using the relations $\boldsymbol{x}\!\cdot\!\boldsymbol{y} = xy\cos{\varphi}$ and  
$\boldsymbol{n}\!\cdot\!\boldsymbol{y} = y\cos{\varphi_y}$, we obtain  
\begin{equation}
  \frac{\boldsymbol{n}\cdot\boldsymbol{x}}{x} = \cos{\varphi_x} =
  \cos{\varphi}\cos{\varphi_y} + \sin{\varphi}\sin{\varphi_y}.
\end{equation}

Employing the integral representation of Bessel functions\,\cite{abramowitz+stegun},
\begin{equation}
  J_n(z) = \frac{i^{-n}}{\pi}%
  \int_0^\pi e^{iz\cos{\varphi}}\cos{(n\varphi)}\,d\varphi,
\end{equation}
the angular integration in Eq.\,\eqref{eq:FourierIntegral} can be performed explicitly for $k = 0$ and $k = 1$:
\begin{equation}
  \int_0^{2\pi} \frac{d\varphi}{2\pi} e^{\pm ixy\cos{\varphi}}
    \begin{cases}
      1 \\
      \cos{\varphi_x}
    \end{cases}
  =
  \begin{cases}
    J_0(xy) \\
    \pm i\cos{\varphi_y}\,J_1(xy)
  \end{cases}.
\label{eq:cos0,1}
\end{equation}

\noindent For higher powers, such as $k = 2$ and $k = 3$, the angular integrals can still be reduced using the results in Eq.\,\eqref{eq:cos0,1}. To achieve this, we neglect terms odd in $\varphi$ (which do not contribute to the amplitudes of interest), and retain only those relevant to the nonflip and spin-flip components. This leads to the following replacements:
\begin{flalign}
  \cos^2{\varphi_x} &\to
  \cos^2{\varphi}\cos^2{\varphi_y} + \sin^2{\varphi}\sin^2{\varphi_y}
  \nonumber \\ &\to
  \frac{1 + \cos{2\varphi_y}}{2}\cos^2{\varphi} + 
  \frac{1 - \cos{2\varphi_y}}{2}\sin^2{\varphi}
  \nonumber \\ &\to
  \frac{1}{2},
  \label{eq:cos2}
  \\
  \cos^3{\varphi_x} &\to
  \cos^3{\varphi_y}\cos^3{\varphi} +
  3(\cos{\varphi_y}\!-\!\cos^3{\varphi_y})\cos{\varphi}\sin^2{\varphi}
  \nonumber \\ &\to
  \frac{3}{4}\cos{\varphi_y}\cos{\varphi}.
  \label{eq:cos3}
\end{flalign}

Integration over $x$ can be performed using the following definite integral\,\cite{Gradshteyn:2007}:
\begin{flalign}
  & \int_0^\infty dx\,x^\mu e^{-\alpha x^2}J_\nu(xy) =  \nonumber \\
  &\quad\frac{y^\nu\, \Gamma\left(\frac{\nu+\mu+1}{2}\right)}%
  {2^{\nu+1}\alpha^{\frac{\nu+\mu+1}{2}}\Gamma(\nu+1)}%
  \,{}_1F_1\left( \frac{\nu+\mu+1}{2};\nu+1; -\frac{y^2}{4\alpha} \right),
  \\
  &\qquad\qquad\left[\mathrm{Re}\,\alpha > 0,\quad%
    \mathrm{Re}(\mu + \nu) > -1\right],
  \nonumber
\end{flalign}
where  
\begin{align}
  {}_1F_1(a;b;z) &= 1 + \frac{a}{b}\frac{z}{1!}%
  + \frac{a(a+1)}{b(b+1)}\frac{z^2}{2!}
  \nonumber \\
  &\quad + \frac{a(a+1)(a+2)}{b(b+1)(b+2)}\frac{z^3}{3!} + \dots
  \label{eq:1F1}
\end{align}
is the confluent hypergeometric function. Notably,
\begin{equation}
   {}_1F_1(a;a;z) = e^z.
\end{equation}
Using the series expansion in Eq.\,\eqref{eq:1F1}, one can show that  
\begin{equation}
   {}_1F_1(1;2;z) = \frac{e^z - 1}{z}.
\end{equation}

In addition to Bessel functions of the first kind and the confluent hypergeometric function, the calculation of Coulomb corrections involves other special functions such as the modified Bessel functions of the first and second kinds, $I_0(z)$ and $K_0(z)$, the gamma function $\Gamma(z)$, and exponential integrals $\mathrm{Ei}(x)$ and $E_1(z)$. Numerical evaluations of these special functions were performed using the GNU Scientific Library (GSL)\,\cite{GSL:2009}.

Some of the definite integrals and expansions used are\,\cite{Gradshteyn:2007,abramowitz+stegun}:
\begin{flalign}
  & \int_0^\infty dx\,\ln{x}\,J_1(ax) = -\frac{1}{a}\left[\ln{\left(\frac{a}{2}\right)} + \gamma\right],
  \label{eq:intM0}
  \\
  & \int_0^\infty dx\,x^\mu J_\nu(xy) = 
  \frac{2^\mu}{y^{\mu+1}}
  \frac{\Gamma\left(\frac{1+\nu+\mu}{2}\right)}%
       {\Gamma\left(\frac{1+\nu-\mu}{2}\right)},
  \label{eq:intM}   \\
  &\qquad\qquad\left[-\mathrm{Re}\,\nu - 1 < \mathrm{Re}\,\mu < \frac{1}{2},\quad y > 0\right],
  \nonumber \\
  & \mathrm{Ei}(x) = -\int_x^\infty \frac{e^{-t}}{t}\,dt, \qquad [x \ne 0],
  \\
  & E_1(z) = \int_z^\infty \frac{e^{-t}}{t}\,dt, \qquad \left|\mathrm{Arg}(z)\right| < \pi,
  \label{eq:E1} \\
  & K_0(ak) = \int_0^\infty \frac{x J_0(ax)}{x^2 + k^2}\,dx, \quad [a\!>\!0,\ \mathrm{Re}(k)\!>\!0].
  \label{eq:K0}
\end{flalign}
\begin{align}
  J_n(z) &=%
  \left(\frac{z}{2}\right)^n\sum_{k=0}^\infty{\frac{(-z^2/4)^k}{k!(k+n)!}},
  \label{eq:sJ0} \\
  I_0(z) &=%
  \sum_{k=0}^\infty \frac{z^{2k}}{(2^k k!)^2},
  \label{eq:sI0} \\
  E_1(z) &= -\gamma - \ln{z} - \sum_{n=1}^\infty \frac{(-1)^n z^n}{n n!},
\end{align}
where $\gamma = 0.5772\dots$ is the Euler-Mascheroni constant.

\section{Fourier Integrals for Amplitudes with Exponential Form Factors
\label{sec:Fourier} }

For exponential form factors (expressed in terms of $q^2$), the eikonal phases can be evaluated analytically or with minimal numerical effort. Using the definite integrals discussed in Appendix\,\ref{sec:Integrals}, we obtain the following expressions.

\noindent\textit{Hadronic nonflip amplitude:}
\begin{align}
  \hat{\gamma}_N(b, B_N) 
  &= \frac{1}{2\pi} \int d^2\boldsymbol{q} \, e^{i\boldsymbol{q} \cdot \boldsymbol{b}} \, e^{-B_Nq^2/2}
  \nonumber \\
  &= \int_0^\infty dq \, q \, e^{-B_Nq^2/2} J_0(qb)
  \nonumber \\
  &= B_N^{-1} \, {}_1\!F_1(1, 1, -b^2/2B_N)
  \nonumber \\
  &= \frac{e^{-b^2/2B_N}}{B_N}.
  \label{eq:fbN}
\end{align}

\noindent\textit{Hadronic spin-flip amplitude:}
\begin{align}
  \hat{\gamma}_S(\boldsymbol{b}, B_S) &=
  \frac{1}{2\pi} \int d^2\boldsymbol{q} \, e^{i\boldsymbol{q} \cdot \boldsymbol{b}} \,
  \cos{\varphi_q} \, q \, e^{-B_Sq^2/2}
  \nonumber \\
  &= i \cos{\varphi_b} \int_0^\infty dq \, q^2 \, e^{-B_Sq^2/2} \, J_1(bq)
  \nonumber \\
  &= i \cos{\varphi_b} \, B_S^{-2} \, {}_1\!F_1(2, 2, -b^2/2B_S)
  \nonumber \\
  &= i \cos{\varphi_b} \, \frac{b \, e^{-b^2/2B_S}}{B_S^2}.
  \label{eq:fbS}
\end{align}

\noindent\textit{Electromagnetic spin-flip amplitude:}
\begin{align}
  \hat{\gamma}_M(\boldsymbol{b}, B_M) &=
  \frac{1}{2\pi} \int d^2\boldsymbol{q} \, e^{i\boldsymbol{q} \cdot \boldsymbol{b}} \,
  \frac{\cos{\varphi_q}}{q} \, e^{-B_Mq^2/2}
  \nonumber \\
  &= i \cos{\varphi_b} \int_0^\infty dq \, e^{-B_Mq^2/2} \, J_1(bq)
  \nonumber \\
  &= i \cos{\varphi_b} \, (2B_M)^{-1} \, {}_1\!F_1(1, 2, -b^2/2B_M)
  \nonumber \\
  &= i \cos{\varphi_b} \, \frac{1 - e^{-b^2/2B_M}}{b}.
\end{align}

\noindent\textit{Electromagnetic nonflip (Coulomb) amplitude:}

\begin{figure}[t]
  \begin{center}
    \includegraphics[width=0.85\columnwidth]{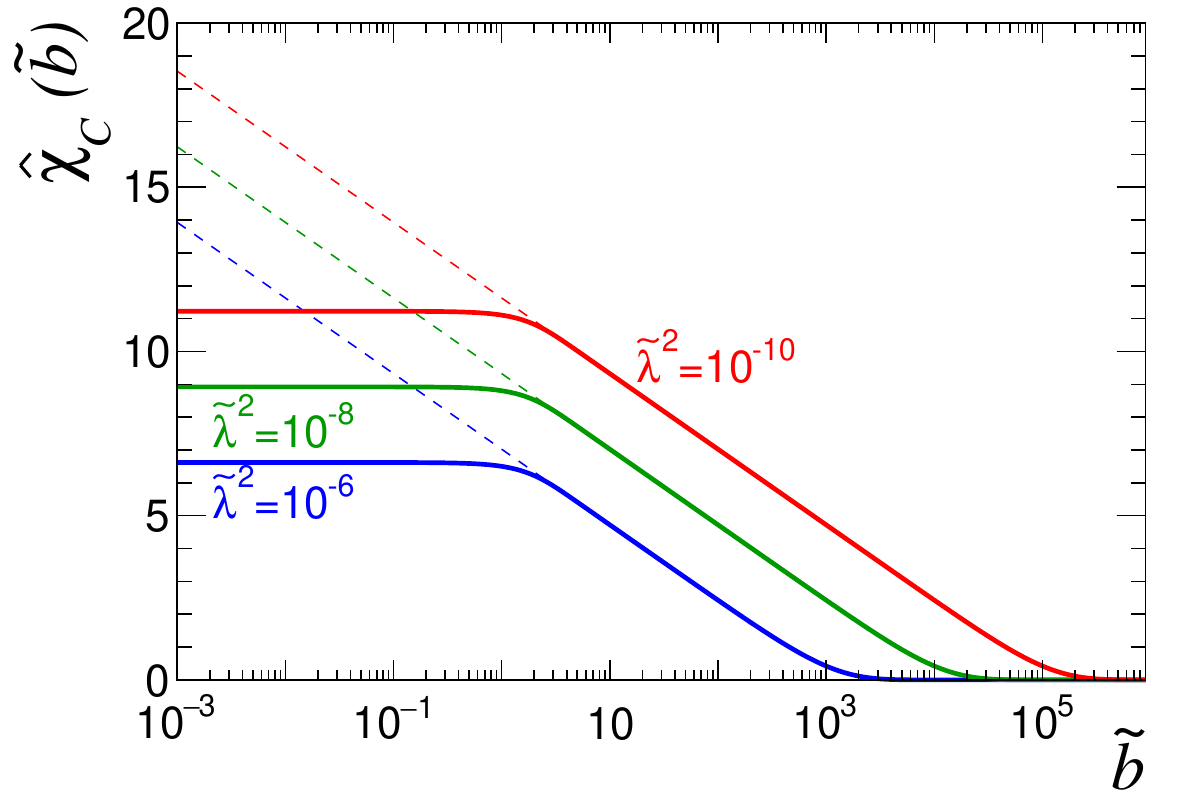}
  \end{center}
  \vspace{-2em}
  \caption{Dependence of the calculated eikonal phase $\hat{\chi}_C(\tilde{b},\tilde{\lambda}^2)$ on the fictitious photon mass. Dashed lines show the corresponding function $\exp(\tilde{\lambda}^2) K_0(\tilde{b} \tilde{\lambda})$.
    \label{fig:fbC}
  }
\end{figure}

Due to the divergence of the Fourier integral for the Coulomb amplitude, numerical integration is performed using a fictitious photon mass $\lambda$:
\begin{align}
  \hat{\chi}_C(\tilde{b}, \tilde{\lambda}^2) &= \tilde{f}_C(\tilde{b}) =
  \frac{1}{2\pi}
  \int d^2\boldsymbol{\tilde{q}} \, e^{i\boldsymbol{\tilde{q}} \cdot \boldsymbol{\tilde{b}}}
  \, \frac{e^{-\tilde{q}^2}}{\tilde{q}^2 + \tilde{\lambda}^2}
  \nonumber \\
  &= \int_0^\infty \frac{\tilde{q} \, d\tilde{q}}{\tilde{q}^2 + \tilde{\lambda}^2}
  \, e^{-\tilde{q}^2/2} \, J_0(\tilde{q} \tilde{b}),
  \label{eq:fbC}
\end{align}
where, using the Coulomb form factor slope $B_C$, the dimensionless variables are defined as
\begin{equation}
  \tilde{q}^2 = \frac{B_C q^2}{2}, \quad
  \tilde{b}^2 = \frac{2 b^2}{B_C}, \quad
  \tilde{\lambda}^2 = \frac{B_C \lambda^2}{2}.
  \label{eq:tilde}
\end{equation}

The dependence of $\hat{\chi}_C(\tilde{b}, \tilde{\lambda}^2)$ on the photon mass is illustrated in Fig.\,\ref{fig:fbC}. For $\tilde{b}\!=\!0$, one can relate integral \eqref{eq:fbC} to the exponential integral \eqref{eq:E1} obtaining
\begin{equation}
  \hat{\chi}_\lambda =\hat{\chi}_C(0,\tilde{\lambda}^2)= 
  \frac{e^{\tilde{\lambda}^2}E_1(\tilde{\lambda}^2)}{2},
  \label{eq:chiL}
\end{equation}
For large $\tilde{b}$, the eikonal phase can be approximated by the Macdonald function \eqref{eq:K0} as
\begin{align}
  \hat{\chi}_C(\tilde{b},\tilde{\lambda}^2) \big|_{\tilde{b} \gtrsim 10} &=
  e^{\tilde{\lambda}^2} K_0(\tilde{\lambda} \tilde{b}).
  \label{eq:chiK0}
\end{align}

\begin{figure}[t]
  \begin{center}
    \includegraphics[width=0.85\columnwidth]{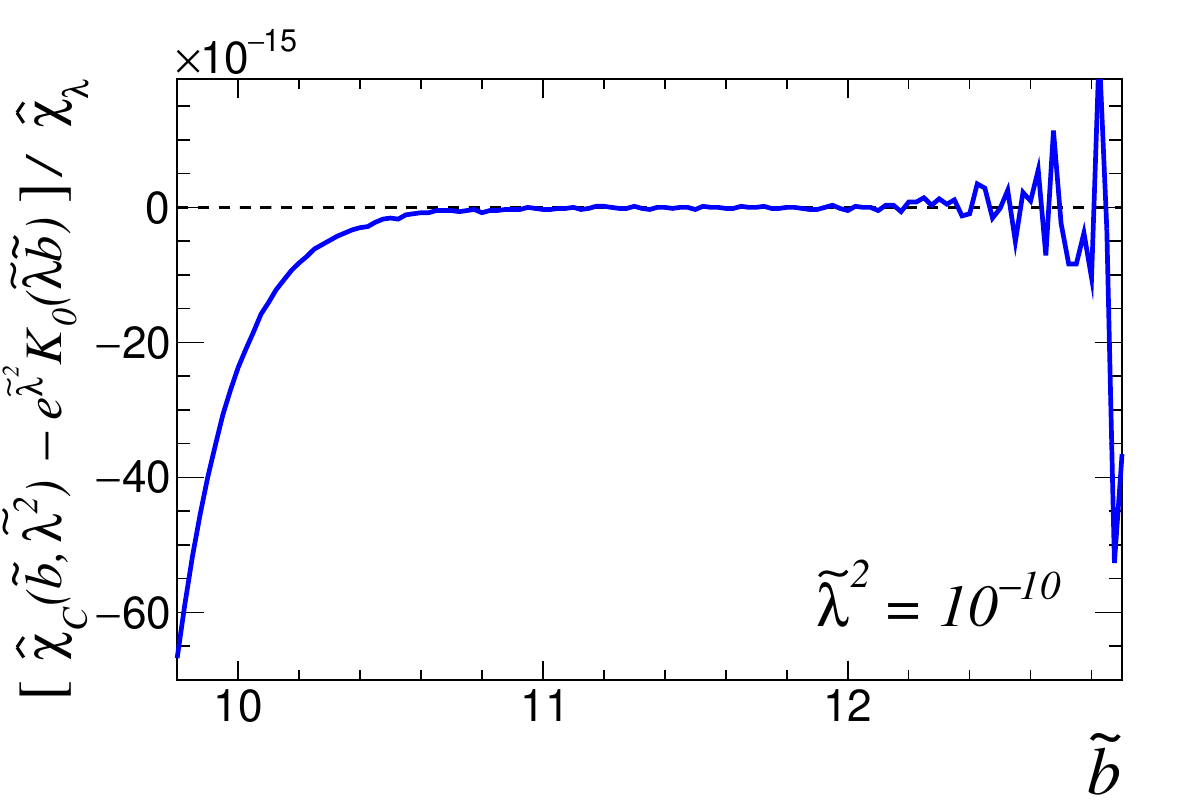}
  \end{center}
  \vspace{-2em}
  \caption{
    \label{fig:fbC-K0}
    Normalized difference between $\hat{\chi}_C(\tilde{b},\tilde{\lambda}^2)$ and $\exp(\tilde{\lambda}^2) K_0(\tilde{b} \tilde{\lambda})$ for $\tilde{\lambda}^2 = 10^{-10}$. Calculation instability for $\tilde{b}\!>\!12$ can be attributed to the computational uncertainties as it was explained in the text.
  }
\end{figure}

Denoting $u\!=\!(\tilde{b}\tilde{q})^2$, $v\!=\!(\tilde{b}\tilde{\lambda})^2$ and using expansions \eqref{eq:sJ0}, \eqref{eq:sI0}, and 
\begin{align}
  \frac{u^n}{u+v} &=%
  \frac{(-v)^n}{u+v}+\sum_{k=0}^{n-1}{(-v)^ku^{n-k-1}},
\end{align}
the Coulomb eikonal phase expression can be transformed to
\begin{flalign}
  &\hat{\chi}_C(\tilde{b}, \tilde{\lambda}^2) = \hat{\chi}_\lambda
  + \int_0^\infty \frac{\tilde{q} \, d\tilde{q}}{\tilde{q}^2 + \tilde{\lambda}^2}
  \, e^{-\tilde{q}^2/2} \left[J_0(\tilde{q}\tilde{b})-1\right], \nonumber \\
  &= \hat{\chi}_\lambda +
  \frac{1}{2}\int_0^\infty{\frac{du}{u+v}e^{-u/{\tilde{b}^2}}\left[I_0(v)-1\right]}%
  \nonumber \\ &
  + \frac{1}{2}\int_0^\infty{du\,e^{-u/{\tilde{b}^2}} \sum_{n=1}^\infty{ \sum_{k=0}^{n-1}{%
        \frac{(-1)^{k+n}v^ku^{n-k-1}}{(2^nn!)^2}
  }}}
\end{flalign}
After integration and regrouping terms we arrive at
\begin{flalign}
  \hat{\chi}_C(\tilde{b},\tilde{\lambda}^2) &=
  \hat{\chi}'_C(\tilde{b}) + \hat{\chi}_0(\tilde{b},\tilde{\lambda}^2)
  + \sum_{n=1}^\infty{\tilde{\lambda}^{2n}\hat{\chi}_n(\tilde{b})}
  \label{eq:chiC}
\end{flalign}
where
\begin{align}
  \hat{\chi}'_C(\tilde{b}) &= -\frac{1}{2}%
  \left[\ln\left(\tilde{b}^2/4\right)%
    + E_1\left(\tilde{b}^2/4\right)\right],
  \label{eq:chi'}
  \\
  \hat{\chi}_0(\tilde{b},\tilde{\lambda}^2) &=
  \hat{\chi}_\lambda I_0(\tilde{b}\tilde{\lambda})-\gamma/2
  \label{eq:chi0}
  \\
  \hat{\chi}_{n\ge1}(\tilde{b}) &=
  \sum_{k=n+1}^\infty{ %
    \frac{(-1)^{k\!+\!n}(k\!-\!n\!-\!1)!}{(k!)^2}\,\left(\frac{b}{2}\right)^{2k} }.
  \label{eq:chi>0}
\end{align}
\
In the double-precision floating point format, commonly used in the numerical calculations, relative accuracy of a rational number is limited by number of bits used for the fraction and can be evaluated as $2^{-53}\!\approx\!10^{-16}$. Since the largest absolute value of a term in series $\hat{\chi}_1$ is achieved at $k\!=\!k_m\!\approx\!\tilde{b}^2/4$, the computational uncertainty of the calculation of $\hat{\chi}_1(\tilde{b})$, estimated by order of magnitude as $\sim2^{-53}\exp{(k_m)}/\sqrt{2\pi k_m}$, is enormous for  large $\tilde{b}$. Nevertheless, for $\tilde{b}\!<\!12$ summation \eqref{eq:chi>0} can be performed with sufficient precision. In particular, for $\tilde{\lambda}^2\!=\!10^{-10}$ and $\tilde{b}\!=\!11$, relative contribution of the series to the eikonal phase can be evaluated as
\begin{align}
  \tilde{\lambda}^2\hat{\chi}_1(\tilde{b},\tilde{\lambda}^2)/\chi_\lambda &=%
  -3.3\times10^{-10}\pm\mathcal{O}(10^{-18}),
  \\
  \tilde{\lambda}^4\hat{\chi}_2(\tilde{b},\tilde{\lambda}^2)/\chi_\lambda &=%
  -1.4\times10^{-19}\pm\mathcal{O}(10^{-29}).
\end{align}

Due to discrete structure of the floating point numbers, summation \eqref{eq:chi>0} is effectively limited by  finite number of terms (e.g., $k<101$ if $\tilde{\lambda}^2\!=\!10^{-10}$).  For large $k$, adding the corresponding term to the already calculated sum does not numerically alter the sum.  

Although the uncertainties very sharply grow (if $\tilde{b}\!>\!12.5$) with increasing of $\tilde{b}$, $\hat{\chi}_C(\tilde{b},\tilde{\lambda}^2)$ can be accurately evaluated for any large $\tilde{b}$ by applying approximation \eqref{eq:chiK0}, as demonstrated in Fig.\,\ref{fig:fbC-K0}. 

Thus, for $\tilde{\lambda}^2\!=\!10^{-10}$, the Coulomb eikonal phase can be calculated as
\begin{flalign}
  \hat{\chi}_C(\tilde{b},\tilde{\lambda}^2) &=
  \begin{cases}
    \hat{\chi}'_C(\tilde{b}) + \hat{\chi}_0(\tilde{b},\tilde{\lambda}^2)
  + \tilde{\lambda}^{2}\hat{\chi}_1(\tilde{b}), & \tilde{b}<11, \\[7pt]
  e^{\tilde{\lambda}^2} K_0(\tilde{\lambda} \tilde{b}), & \tilde{b}\ge11.
  \end{cases}
  \label{eq:chiC-10}
\end{flalign}
For  $\tilde{\lambda}^2\!>\!10^{-8}$, term $\tilde{\lambda}^{4}\hat{\chi}_2(\tilde{b})$
may also be needed to consider (if $\tilde{b}\!<\!11$).

\section{Modified Eikonal Phase for the Coulomb Amplitude
\label{sec:chiC'}}

A typical expression for the Coulomb-corrected amplitude $f_X^\gamma(q^2)$ is given by
\begin{align}
  f_X^\gamma(q^2) &= \int_0^\infty b\,db\,\gamma_X(b)\,
  e^{i\chi_C(b, \lambda^2)} J_{0,1}(bq) \nonumber \\
  &= f_X(q^2)\mathcal{F}_X(q^2) \approx f_X(q^2)\,e^{i\Phi_X(q^2,\lambda^2)},
  \label{eq:fX}
\end{align}
where $X$ denotes one of the indices $N$, $S$, or $M$. The Bessel function $J_0(bq)$ is used for the nonflip amplitude ($X = N$), while $J_1(bq)$ is used for the spin-flip amplitudes ($X = S, M$). The eikonal phase $\chi_C(b, \lambda^2)$ is derived from the reduced phase \eqref{eq:chiC-10} by multiplying by the Coulomb coupling:
\begin{equation}
  \chi_C(b,\lambda^2) = -2\alpha Z\,%
  \hat{\chi}_C\left(b\sqrt{2/B_C},\lambda\sqrt{B_C/2}\right).
\end{equation}

Since adding a constant to $\chi_C(b,\lambda^2)$ uniformly shifts all Coulomb phases $\Phi_X$, the nonvanishing $\lambda^2$ dependence of $\Phi_X(q^2,\lambda^2)$ as $\lambda \to 0$ can be eliminated by subtracting $\hat{\chi}_C(0,\tilde{\lambda}^2)$ from $\hat{\chi}_C(\tilde{b},\tilde{\lambda}^2)$ \cite{Poblaguev:2024yui}. Consequently, replacing $\tilde{\lambda}\!\to0$ in the corrected phase, we obtain the modified eikonal phase in the massless photon limit:
\begin{align}
  \chi'_C(b) &= -2\alpha Z \hat{\chi}'_C(\tilde{b})
  \nonumber \\
  &= \alpha Z\left[\ln\left(\frac{b^2}{2B_C}\right) + E_1\left(\frac{b^2}{2B_C}\right)\right].
  \label{eq:chiC'}
\end{align}

\begin{figure}[t]
  \begin{center}
    \includegraphics[width=0.85\columnwidth]{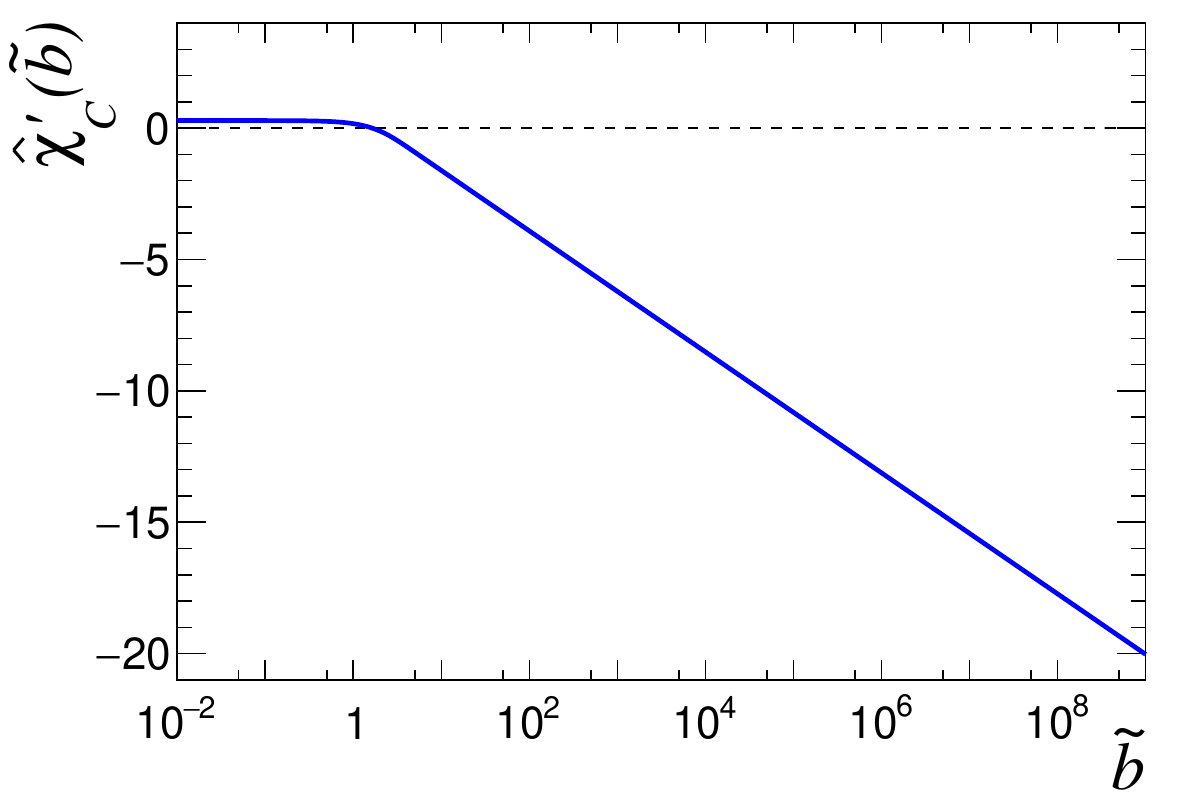}
  \end{center}
  \vspace{-2em}
  \caption{The modified Couloumb eikonal phase (as a function of the dimenshionless $\tilde{q}^2$). $\hat{\chi}'_C(\tilde{b})\!\to\!\gamma/2$ if $\tilde{b}\!\to\!0$, and $\hat{\chi}'_C(\tilde{b})\!\to\!\ln{(\tilde{b}/2)}$ for large $\tilde{b}$.
    \label{fig:chiC0}
  }
\end{figure}

Using this modified phase (shown in Fig.\,\ref{fig:chiC0} requires a corresponding adjustment of the Coulomb correction to the nonflip electromagnetic amplitude:
\begin{equation}
  f_C^\gamma(q^2,\lambda^2) = e^{-i\Delta} \int_0^\infty b\,db\,
  i\left[1 - e^{i\chi_C(b, \lambda^2)}\right] J_0(bq),
  \label{eq:fC}
\end{equation}
with
\begin{equation}
  \Delta = \alpha Z \left[ e^{B_C \lambda^2 / 2} E_1\left(\frac{B_C \lambda^2}{2}\right) + \gamma \right].
  \label{eq:Delta}
\end{equation}

Although this shift cancels $\log{\lambda^2}$ terms in the calculated phase $\Phi_C(q^2,\lambda^2)$, it does not completely eliminate the dependence on the fictitious photon mass $\lambda$. However, this dependence vanishes smoothly as $\lambda \to 0$.

\section{Coulomb Corrections to the Electromagnetic Spin-Flip Amplitudes
\label{sec:FgM}}

The Coulomb correction factor $\mathcal{F}_M(q^2)$ for the electromagnetic spin-flip amplitude, defined in accordance with Eq.\,\eqref{eq:fX},
\begin{align}
  f_M^\gamma(q^2) &= \int_0^\infty b\,db\,\gamma_M(b)\,
  e^{i\chi'_C(b)} J_{1}(bq) = f_M(q^2)\mathcal{F}_M(q^2) \nonumber \\
  &= f_M(q^2) \times \widetilde{\mathcal{F}}(B_C q^2/2,B_C/B_M),
  \label{eq:FC}
\end{align}
 can be straightforwardly expressed in terms of a function of the dimensionless variables $\tilde{q}^2\!=\!B_Cq^2/2$ and $\tilde{\beta}\!=\!B_C/B_M$: 
\begin{flalign}
  &\widetilde{\mathcal{F}}(\tilde{q}^2,\tilde{\beta}) = \nonumber \\
  &\quad\tilde{q}\,e^{\tilde{q}^2}%
  \int_0^\infty\!\!db\,\left(1\!-\!e^{-\frac{\tilde{\beta}b^2}{4}}\right)%
  e^{i\alpha{Z}\left[\ln{\frac{b^2}{4}}\!+\!E_1\left(\frac{b^2}{4}\right)\right]} J_1(b\tilde{q}).
\end{flalign}

For $b > 20$, both $\exp{(-b^2/4)}$ and $E_1(b^2/4)$ become negligible (less than $10^{-45}$), and the integrand effectively reduces to that in the definite integral \eqref{eq:intM}. Thus, the Coulomb correction can be well-approximated by the expression
\begin{flalign}
  &\widetilde{\mathcal{F}}(\tilde{q}^2,\tilde{\beta}) =%
  e^{\tilde{q}^2} \times  \bigg\{
  e^{-i\alpha{Z}\ln{q^2}}\,%
  \frac{\Gamma\left(1 + i\alpha{Z}\right)}{\Gamma\left(1 - i\alpha{Z}\right)}
  - I_R(20\tilde{q})
  \nonumber \\&
  + \tilde{q} \int_0^{20} \!\! db \, e^{i\alpha{Z}\left[\ln{\frac{b^2}{4}} + E_1(\frac{b^2}{4})\right]}%
  \left(1\!-\!e^{-\frac{\tilde{\beta}b^2}{4}}\right) J_1(\tilde{q}b)
  \bigg\},
  \label{eq:FM}
\end{flalign}
where
\begin{equation}
  I_R(a) = \tilde{q} \int_0^{a/\tilde{q}}{\!\! db \,
    e^{i\alpha{Z}\ln{(b^2/4)}} J_1(\tilde{q}b). }
\end{equation}
Denoting $t\!=\!b\tilde{q}$ and expanding $J_1(t)$, one can substitute $I_R(a)$ by the sum
\begin{equation}
  I_R(a)\to\Sigma_R(a)=%
  \frac{a^2}{4}%
  \sum_{k=0}^\infty{\frac{(-a^2/4)^k}{k!(k+1)!}\,%
    \frac{ e^{i\alpha{Z}\ln{\frac{a^2}{4\tilde{q}^2}} }}{k+1+i\alpha{Z}}
  }.
  \label{eq:IR}
\end{equation}  

The series \eqref{eq:IR} converges rapidly for $a\!<\!2$, allowing it to be evaluated with high accuracy, limited only by computational discretization. As a result, the integral $I_R(20\tilde{q})$ can be partially replaced by a fast converging summation:  
\begin{equation}
  I_R(20\tilde{q}) = 
  \begin{cases}
    \Sigma_R(20\tilde{q}),  & \tilde{q}\leq0.1\\[7pt]
    \Sigma_R(2)+\int_2^{20\tilde{q}}{\!dt\,%
      e^{i\alpha{Z}\ln{\frac{t^2}{4\tilde{q}^2}}}
        J_1(t)},
      &\tilde{q}>0.1
  \end{cases}
  \label{eq:IRcases}
\end{equation}

Numerical integration of the explicitly written integrals in  Eqs.\,\eqref{eq:FM} and \eqref{eq:IRcases} can be done using equidistance sampling points and, thus, it does not pose any significant difficulties.

The spin-flip phase in the leading-order $\alpha$ approximation (allowing direct comparison with the analytical result from Ref.\,\cite{Kopeliovich:2000ez} for the nonflip electromagnetic amplitude) can be evaluated numerically in a similar way. Defining the phase as in Ref.\,\cite{Kopeliovich:2000ez} and using the definite integral \eqref{eq:intM0}, we obtain
\begin{flalign}
  &\Phi_M^\text{LO}(\tilde{q}^2) = \frac{\int b\,db\,\hat{\chi}_M(b)\chi'_C(b)}{\int b\,db\,\hat{\chi}_M(b)}
  \nonumber \\
  &~= -2\alpha{Z} e^{\tilde{q}^2} \times \bigg\{ 
    \ln{\tilde{q}} + \gamma \nonumber
  + \tilde{q} \int_0^{20} db\,\ln{(b/2)} J_1(b\tilde{q}) 
  \nonumber \\
  &~- \frac{\tilde{q}}{2} \int_0^{20} \!\!db 
  \left(1\!-\!e^{-\frac{b^2}{4}}\right) 
  \left[\ln{\frac{b^2}{4}}\!+\!E_1(\frac{b^2}{4})\right] J_1(b\tilde{q})
  \bigg\}.
  \label{eq:PhiM_LO}
\end{flalign}

\section*{Acknowledgments}
This work is authored by employee of Brookhaven Science Associates, LLC under Contract No. DE-SC0012704 with the U.S. Department of Energy.

\section*{DATA AVAILABILITY}

The data are not publicly available. The data are available from the authors upon reasonable request.

\bibliographystyle{apsrev4-2}
\input{v1.4_NumericalCoulomb_arXiv.bbl}

\end{document}

%% file: v1.4_NumericalCoulomb_arXiv.bbl
%